\documentclass[journal]{IEEEtran}

\usepackage{subfigure}
\usepackage{amsmath,amssymb}
\usepackage[dvips]{graphicx}
\usepackage{amsfonts}
\usepackage[mathscr]{eucal}
\usepackage{latexsym}
\usepackage{amsthm}
\usepackage{exscale}
\usepackage[mathscr]{eucal}
\usepackage{bm}
\usepackage[dvipsnames]{color}
\usepackage{cases}
\usepackage{epsfig}
\usepackage[center,small]{caption}
\usepackage{algorithm}
\usepackage{algorithmic}
\usepackage[verbose,nospace,sort]{cite}
\usepackage{tabularx,ragged2e}
\usepackage{multirow}
\usepackage{multicol}
\usepackage{balance}
\usepackage{array}

\scrollmode

\newcommand\mycom[2]{\genfrac{}{}{0pt}{}{#1}{#2}}

\newcommand{\tr}{\mathrm{tr}}

\newcommand{\rf}{\mathrm{rf}}
\newcommand{\dc}{\mathrm{dc}}
\newcommand{\jImag}{\jmath}

\graphicspath{{./figs/}}

\begin{document}
\title{Communications and Signals Design for Wireless Power Transmission}

\author{\authorblockN{Yong~Zeng,  Bruno Clerckx, and Rui~Zhang}
\thanks{Y. Zeng and R. Zhang are with the Department of Electrical and Computer Engineering, National University of Singapore, Singapore 117583 (e-mail: \{elezeng, elezhang\}@nus.edu.sg). R. Zhang is also with the Institute for Infocomm Research, A*STAR, Singapore 138632.}
\thanks{B. Clerckx is with the EEE department at Imperial College London, London SW7 2AZ, United Kingdom (email: b.clerckx@imperial.ac.uk).}
}
\IEEEspecialpapernotice{(Invited Paper)}

\maketitle


\begin{abstract}
Radiative wireless power transfer (WPT) is a promising technology to provide cost-effective and real-time power supplies to wireless devices. Although radiative WPT shares many similar characteristics  with the extensively studied wireless information transfer or communication, they also differ significantly in terms of  design objectives, transmitter/receiver architectures and hardware constraints, etc. In this article, we first give an overview on the various WPT technologies, the historical development of the radiative WPT technology and the main challenges  in designing contemporary radiative WPT systems. Then, we focus on discussing the new communication and signal processing techniques that can be applied to tackle these challenges. Topics discussed include energy harvester modeling, energy beamforming for WPT, channel acquisition, power region characterization in multi-user WPT, waveform design with linear and non-linear energy receiver model, safety and health issues of WPT, massive MIMO (multiple-input multiple-output) and millimeter wave (mmWave) enabled WPT, wireless charging control, and wireless power and communication systems co-design. We also point out directions that are promising for future research.

\end{abstract}


\section{Introduction}
Traditionally, electronic devices such as cell phones, laptops, digital cameras, etc. are mostly powered by batteries, which have limited energy storage capacity and thus need to be regularly recharged  or replaced. With the widespread use of portable electronic devices during the past decade, mainly driven by the fast growing market on smart phones, tablets, wearable electronic devices, etc., there is also an ever-increasing interest for powering devices wirelessly. Compared to the conventional battery, wireless charging is a promising alternative that is in general more user-friendly by eliminating the hassle of connecting cables, more cost-effective by enabling on-demand energy supplies and uninterrupted operations, more environmental preserving  by avoiding massive battery disposal, and sometimes  essential for applications in which manual battery replacement/recharging is dangerous (e.g., in hazardous environment) or even impossible (e.g., for biomedical implants). The key enabler for wireless charging is the advancement of dedicated wireless power transfer (WPT) technology \cite{678,670,677,715,716,751}, a collective term that refers to any method of delivering power from one place to another without interconnecting wires. Various WPT technologies have been developed so far,  including {\it inductive coupling, magnetic resonant coupling, electromagnetic (EM) radiation, and laser power beaming}, among others. An overview of them is given in the following.

\begin{table*}
\centering
\caption{Comparison of the main technologies for WPT.}\label{table:comparison}
\begin{tabular}{|p{1.5cm}|>{\raggedright}p{1.5cm}|>{\raggedright}p{1.5cm}|>{\raggedright}p{1cm}|>{\raggedright}p{4cm}|>{\raggedright}p{4cm}|p{1.8cm}|}
\hline
{\bf WPT technology}& {\bf Main devices} & {\bf Typical range} &{\bf Typical frequency} & {\bf Main advantages and limitations} & {\bf Current and potential applications} & {\bf Representative companies}\\
\hline
{\bf Inductive coupling} & Wire coils & Millimeters to centimeters & Hz to MHz &
 High efficiency, require precise tx/rx coil alignment, very short range, single receiver only & Electric tooth brush and razor battery charging, transcutaneous charging of bio-medical implants, electrical vehicle charging, cell phone charging, factory automation & Powermat, Delphi, GetPowerPad, WildCharge, Primove\\
\hline
{\bf Magnetic resonant coupling} & Tuned wire coils, lumped element resonators & A few meters, typically 4 to 10 times the coil diameter & kHz to MHz & High efficiency, safe, mid-range, large tx/rx size & Consumer electronics (e.g., cell phones, laptops, household robots) charging, biomedical implants charging, electrical vehicles charging, RFID, smart cards, industrial applications & PowerbyProxi, WiTricity, WiPower, Intel (Wireless Resonant Energy Link)  \\
\hline
{\bf EM radiation} & Dish antenna, antenna array, rectenna & Several meters to hundreds of kilometers & MHz to dozens of GHz & Long range, small receiver form factors, flexible in deployment and movement, support power multicasting, potential for SWIPT, LoS link is not a must, low efficiency, safety and health issues & Wireless sensor charging, IoT, RFID, consumer electronics charging, wireless-powered aircrafts, solar power satellite & Intel (WISP), Energous (Wattup),  PowerCast, Ossia (Cota) \\
\hline
{\bf Laser power beaming} & Laser emitter, photovoltaic receiver & up to kilometers & THz & Compact size, high energy concentration, no interference to existing communication systems or electronics, laser radiation is hazardous, require LoS link and accurate receiver targeting, vulnerable to atmospheric absorption and scattering by clouds, fog, and rain &  Laser-powered UAVs, laser-powered space elevator climbers, laser-based solar power satellite & LaserMotive \\
\hline
\end{tabular}
\end{table*}

\subsection{Overview of WPT Technologies}
{\it Inductive coupling} is a near-field WPT technology where power is transferred between two properly aligned transmitter/receiver coils by magnetic field \cite{674,675,673,676,672,671,732}. Similar to transformers, the fundamental principles of inductive WPT are Ampere's law and Faraday's law of induction. The alternating current passing through the transmitter coil creates a time-varying magnetic field, which, upon passing through the receiving coil, induces an alternating current in the receiving circuit that could be converted to usage energy. Inductive coupling is able to achieve high power transfer efficiency (e.g., up to 90\%), but the transmitter and receiver need to be in close proximity and aligned accurately. Thus, inductive coupling is not suitable for  charging multiple devices concurrently when the devices are freely placed in an area.

{\it Magnetic resonant coupling} is another near-field WPT technology  that makes use of the well known principle of resonant coupling \cite{666,501,660}, i.e., two objects resonant at the same frequency tend to couple with each other most efficiently. Though both use magnetic field as the medium for WPT, magnetic resonant coupling is able to achieve higher power transfer efficiency over longer distances than inductive coupling, by carefully tuning the transmitter and receiver  circuits to make them resonant at the same frequency. Furthermore, compared to inductive coupling, WPT via magnetic resonant coupling has a relatively loose requirement on coil alignment. Leveraging this technique, a team from MIT has demonstrated lighting up a 60W light-bulb over 2 meters with about 40\% efficiency \cite{501}, which has since spurred numerous research interests on this topic \cite{670},\cite{662,721,663,664,665,661,669,731,720,730,840,854,855}. Today, several interface standards have been developed for the two near-field WPT technologies, including Qi (pronounced as ``Chee'', coming from the Chinese word meaning ``natural energy'') by the Wireless Power Consortium \cite{667}, and AirFuel by the AirFuel Alliance (a merge of the former Alliance for the Wireless Power and Power Matters Alliance)\cite{668}. Commercial products that support the near-field wireless charging standards are already available in the market.  

{\it EM radiation}, which has been primarily used for wireless communication, is another promising approach for WPT, also known as radiative WPT. In contrast to the two near-field wireless charging methods, radiative WPT is a far-field wireless power transmission technology with the transmitter  and receiver completely decoupled electrically, i.e., the energy absorption by the receiver does not affect the power radiation of the transmitter. In radiative WPT, the modulated/unmodulated energy-bearing signals at the transmitter are up-converted into the designated radio frequency, radiated by the transmitting antennas (e.g., parabolic dish antennas or antenna arrays), propagating through the wireless channel, then picked up by the receiving antennas, and finally converted into the usable direct current (DC) via devices such as rectifiers. Note that the simplest rectifiers usually consist of a matching circuit, a diode, and a low-pass filter \cite{694},\cite{697}. The combination of the energy receiving antenna and the rectifier is termed {\it rectenna} \cite{695,696,756}. Depending on the antenna size, transmitting power, as well as the propagation environment, radiative WPT may achieve power delivery over distances varying from a few meters to even hundreds of kilometers \cite{684}. Besides longer transmission distance, radiative WPT also enjoys many other promising advantages as compared to the near-field WPT counterparts, such as smaller transmitter/receiver form factors, more flexible in transmitter/receiver deployment and movement, more suitable for concurrent power delivery to multiple receivers (i.e., power multi-casting),  applicable even in non-line of sight (NLoS) environment, as well as the potential for  simultaneous wireless information and power transfer (SWIPT) \cite{503,504,478} and wireless powered communications \cite{515,525,744,753,742,743}. As a result, radiative WPT has a wide range of applications, spanning from low-power wireless charging  for devices such as radio frequency identification (RFID) tags, wireless sensors, Internet of Things (IoT) devices, and consumer electronics (smart phones, laptops, household robots, etc.), to high-power applications such as microwave-powered aircrafts \cite{693,506,713,714} as well as solar power satellite (SPS) \cite{698},\cite{682}.  Encouragingly, several startup companies such as Energous (Wattup) \cite{687} and Ossia (Cota) \cite{688} have experimentally demonstrated the feasibility of wirelessly charging smart phones using radiative WPT technology in room-size distance (e.g., 9 meters), which could bring a revolutionizing transform of future generation consumer electronics.

Last but not least, another potential  technology for WPT is {\it laser power beaming}, which uses highly concentrated laser light aiming at the energy receiver to achieve efficient power delivery over long distances \cite{757,758,759}. Similar to solar power, the receiver of laser powering uses specialized photovoltaic cells to convert the received laser light into electricity. One promising application of laser-based WPT technology is to provide essentially perpetual power supply to unmanned aerial vehicles (UAVs) in flight, enabling them potentially unlimited endurance aloft: a vision which would bring drastic performance improvement for numerous UAV-enabled applications \cite{616},\cite{649}. A series of flight tests have been performed by LaserMotive company that successfully demonstrated the great potential of laser-powered UAVs \cite{760}. 
However, laser-based WPT has several limitations. First of all, laser radiation could be hazardous. Secondly, laser beaming requires LoS link as well as accurate pointing towards the receiver, which could be challenging to achieve in practice. Moreover, compared to radiative WPT, laser beaming is more vulnerable to atmospheric absorption and scattering by clouds, fog, and rain, which greatly hinders its practical applications.


Besides dedicated WPT, another promising tetherless  power solution is via passive energy scavenging, where the devices opportunistically harvest the available energy in the surrounding environment that is not intended for power delivery. The viable energy sources that could be harvested include solar, wind, vibration, ambient radio frequency (RF) signals, etc \cite{762,763,764,765,637}. 
Though providing a viable solution for green energy at essentially no overhead energy cost, energy scavenging is subject to various factors that are usually beyond the operator's control, such as weather, transmission power of the surrounding RF transmitters, etc. In contrast, by using dedicated power transmitters, WPT is able to offer stable and fully controllable power supplies to wireless devices of different energy demands, and thus is anticipated to play an important role in future wireless systems.

The comparison of the various WPT technologies above is summarized in Table~\ref{table:comparison}. The rest of this article will be focused on radiative WPT technology considering its great potential for more diversified applications compared to other alternatives.

\subsection{History of Radiative Wireless Power Transfer}
The history of WPT by radio waves can be traced back to the early work by Heinrich Hertz in 1880's \cite{689}, whose purpose was to demonstrate the existence and propagation of electromagnetic waves in free space.  In his experiment, Hertz used a spark-gap transmitter (equivalently a dipole antenna) to generate high-frequency power and detected it at the receiving end, which resembled a complete WPT system. Some years later in 1899, 
Nicola Tesla conducted the first experiment on dedicated power transmission without using wires \cite{691,692,690}. In his experiment, Tesla built a gigantic coil, which was fed with 300 kW power resonating at 150 kHz. However, there was no clear record on whether any significant amount of power was collected at certain point. Thereafter, Tesla started the ambitious Wardenclyffe Tower project in 1901, where a large wireless transmission station  was constructed for transmitting messages, telephony, and wireless power \cite{692}. However, the project was not completed since Tesla failed to get continuous financial support.

During the first half of the 20th century, research on WPT was almost dormant and little progress was made.  With the great advancement of microwave technology during World War II, such as the development of magnetron tubes for high-power microwave generations and more advanced parabolic antennas for highly directional radiations, it was realized that efficient WPT became more feasible and thus the interest on WPT was revived. 
 In 1964, William C. Brown, the pioneer of modern radiative WPT technology, successfully demonstrated a wireless-powered helicopter after the invention of rectenna \cite{693},\cite{506}. In this demonstration, the  helicopter was tethered for the purpose of lateral positioning, flying about 18 meters above the transmitting antenna with all the power (about 270W) received via a 2.45 GHz microwave beam. In 1968, William C. Brown demonstrated a beam-positioned helicopter that uses microwave beam to automatically position the helicopter over the beam center. However, instead of powering by radiative WPT, the helicopter in this demonstration was powered via an umbilical cable. Unfortunately, due to financial issues, no further activity was performed to demonstrate the more interesting system of a completely untethered helicopter that is both powered and positioned by microwave beam \cite{680}.

In 1968, Peter Glaser proposed the SPS concept \cite{698}, which has since profoundly affected the research direction of radiative WPT. The main idea of SPS is to collect the solar energy by a geostationary satellite, convert it into microwave signals, and then transmit to the Earth for use via microwave beam. Due to the ample and more stable solar energy available in  geostationary orbit than at ground, SPS was regarded as an effective approach to solve  the energy shortage and greenhouse gases emission problems, and attracted significant research interests for more than half a century \cite{683,700,699,681,507,719,750}. In 1975, a WPT experiment with an overall DC to DC power transfer efficiency of 54\% is achieved in Raytheon Laboratory, with the transmit and receive antenna separated by 1.7m and the DC output power of 495W \cite{703}. This is the highest radiative WPT efficiency known to date.  In 1975, another remarkable experiment on radiative WPT, known as the JPL (Jet Propulsion Laboratory) Goldstone demonstration, was conducted by William C. Brown and his colleagues \cite{706},\cite{707}. In this experiment, over 30kW of DC power was obtained from the rectenna receiver that was 1.54km away from the transmitter using microwave beam at 2.388GHz, which strongly demonstrated the feasibility of high power transmission over long distance via microwave. This achievement was mainly attributed to three factors: the high transmission power (450kW), the highly efficient rectenna used (with the microwave to DC conversion efficiency of 84\%) \cite{680},\cite{505}, as well as the large transmit and receive antennas employed (26m-diameter dish transmit antenna and a 7.3 $\times$ 3.5 m rectenna array). Such encouraging results led to a comprehensive study of the SPS concept by NASA and the U.S. Department of Energy (DOE), covering technical, environmental and societal aspects, which was completed in 1980. Despite of the favorable conclusion on the SPS concept, it was recommended that the development and deployment of the SPS system should not proceed before the technology became sufficiently mature \cite{704},\cite{705}. Since then, the research on SPS was mostly shifted to Japan.

In 1983, Japan launched the first rocket experiment to test high-power microwave transmission through the ionosphere, known as MINIX project (Microwave Ionosphere Nonlinear Interaction eXperiment). The MINIX experiment demonstrated the power transmission from a daughter vehicle to a mother vehicle in space using a 2.45GHz microwave beam \cite{679,709,708,712,710,711}. 
 In 1987, Canada demonstrated the first free-flying wireless-powered aircraft in the program known as Stationary High Altitude Relay Platform (SHARP) \cite{713},\cite{714}, which was proposed to provide long-endurance low-cost aerial communication relaying platform. In SHARP demonstration, a 2.45-GHz microwave beam was transmitted by a parabolic dish antenna to power the aircraft 150m above the ground level. In 1992, Japan conducted the MILAX (Microwave Lifted Airplane eXperiment) experiment \cite{679},\cite{711} which was the first experiment to apply the electronically steerable phased array transmitter for radiative power transmission. In this experiment, a 2.411-GHz continuous wave (CW) unmodulated signal of power 1.25kW was transmitted by 288-element transmitting array, which was assembled on the roof of a car to move underneath the fuel-free aircraft. 
  At the receiver side, the airplane flying at approximately 10m above the ground level was equipped with a receiving array with 120 rectennas. The maximum DC power obtained from the rectenna array was approximately 88W. In 1993, Japan conducted the ISY-METS (International Space Year-Microwave Energy Transmission in Space) experiment to demonstrate the space to space radiative power transmission \cite{717},\cite{718}. In 1995, an experiment called ETHER (the Energy Transmission toward High-altitude long endurance airship ExpeRiment) was conducted in Japan \cite{711}, which transmitted 2.45-GHz, 10-kW power to an airship flying around 40m above the ground level using parabolic antenna. In 1997, France started the project aiming to deliver 10kW of electricity power wirelessly in the La Reunion island \cite{686}.  In 2008, power was successfully transmitted wirelessly between two islands in Hawaii over 148km \cite{684}. Although only 20W of power was received, the power delivery range in the Hawaii demonstration was significantly larger than prior experiments. In 2015, Japan announced that they successfully beamed 1.8kW power with pinpoint accuracy to a small receiver device 55m away \cite{685}.

The main historical milestones for radiative WPT are summarized in Table~\ref{table:mileStones} in chronological order.

\begin{table}
\centering
\caption{Historical milestones for radiative WPT.}\label{table:mileStones}
\begin{tabular}{|p{0.8cm}|p{6.5cm}|}
\hline
{\bf Year} & {\bf Main activity and achievement} \\
\hline
1888 & Heinrich Hertz demonstrated electromagnetic wave propagation in free space.\\
\hline
1899 & Nicola Tesla conducted the first experiment on dedicated WPT.\\
\hline
1901 & Nicola Tesla started the Wardenclyffe Tower project.\\
\hline
1964 & William C. Brown invented rectenna.\\
\hline
1964 & William C. Brown successfully demonstrated the wireless-powered tethered helicopter.\\
\hline
1968 & William C. Brown demonstrated the beam-positioned helicopter. \\
\hline
1968 & Peter Glaser proposed the SPS concept.\\
\hline
1975 & An overall  DC to DC power transfer efficiency of 54\% was achieved in Raytheon Laboratory.\\
\hline
1975 & Over 30kW DC power was obtained over 1.54km in the JPL Goldstone demonstration. \\
\hline
1983 & Japan launched the MINIX project.\\
\hline
1987 & Canada demonstrated the free-flying wireless-powered aircraft 150m above the ground.\\
\hline
1992 & Japan conducted the MILAX experiment with the phased array transmitter.\\
\hline
1993 & Japan conducted the ISY-METS experiment.\\
\hline
1995 & Japan conducted the ETHER experiment for wireless powering the airship.\\
\hline
1997 & France conducted the Reunion Island project to transmit 10kW power to a remote village.\\
\hline
2008 & Power was successfully transmitted over 148km in Hawaii.\\
\hline
2015 & Japan announced successful power beaming to a small device.\\
\hline
\end{tabular}
\end{table}

\subsection{Radiative Wireless Power Transfer: A Fresh New Look}
As reviewed in the preceding subsection, radiative WPT has been historically targeting for long-distance and high-power transmissions, as mainly driven by the two appealing applications: wireless-powered aircraft and SPS. This usually requires very high transmit power (e.g., 450kW for the JPL Goldstone demonstration), huge transmit  and receive antennas (e.g., 26-m diameter parabolic dish), as well as a clear LoS link between the transmitter and receiver. More recently, there has been a significant interest in radiative WPT for relatively low-power (e.g., from micro-watts to a few watts) delivery over moderate distances (e.g., from a few meters to possibly hundreds of meters) \cite{736}, \cite{735}, owing to the fast-growing need to build reliable and convenient WPT systems for remotely charging various low- to medium-power devices, such as RFID tags \cite{739},\cite{733}, wireless sensors \cite{869,740,502,737,490,738}, consumer electronics including smart phones \cite{741}. Though much lower power needs to be delivered as compared to the ambitious wireless-powered aircraft and SPS applications, future WPT systems that are suitable for daily use are facing many new design challenges, such as more compact transmitter/receiver equipment, more complicated propagation environment, the need to support mobility, the safety and health issues, the potential impact on wireless communication systems, etc. More specifically, the following are the authors' views on the important engineering requirements as well as the main design challenges for future radiative WPT systems.

\subsubsection {\it Range} Depending on the power requirement and receiver sensitivity, future WPT systems are expected to achieve power delivery for distances from a few meters (e.g., for smart phone charging) to hundreds of meters (e.g., for wireless sensor charging).
\subsubsection {\it Efficiency} The end-to-end power transfer efficiency is of paramount importance, and also one of the most challenging design aspects for radiative WPT systems. An effective radiative WPT system is expected to achieve an overall efficiency from a fractional of percent to a few percent, depending on the distance. This requires efficient DC to RF power conversion at the transmitter, highly directive RF transmission or energy beamforming over the air, as well as highly efficient RF to DC conversion at the receiver. For further improved efficiency, an end-to-end design with jointly optimized transmitter and rectennas may need to be pursued.
\subsubsection {\it Non-line of sight} Although LoS is always preferred for efficient power delivery, the ability to support NLoS power transmission would significantly widen the practical applications of future WPT systems, and thus is of high practical interests. Energy beamforming over NLoS environment requires a reasonable power balance along different propagation paths, rather than focusing on a single beaming direction as in LoS scenario. To this end, a closed-loop WPT operation is needed in general, i.e., a reverse communication link from the receiver to the transmitter is used to support various functions such as channel feedback/training, energy feedback, charging control, etc.
\subsubsection {\it Mobility  support} Effective radiative WPT systems need to support power delivery even for moving receivers, at least for those at the pedestrian speed. To this end, the transmitter should be able to flexibly adjust the beam directions, and thus renders the electronically steerable phase array or even the more advanced MIMO (multiple-input multiple-output) technique an indispensable part for radiative WPT systems. This is in a sharp contrast to early radiative WPT designs for static applications, which usually make use of high-aperture parabolic dish antennas but require mechanical antenna adjustment for direction control.
\subsubsection {\it Ubiquitous and authenticated accessibility} Similar to the well-established wireless communication systems, effective WPT systems need to support ubiquitous power accessibility at any location within the power coverage area. This in general requires densely deployed and well coordinated multiple energy transmitters to form a radiative WPT network for cooperative WPT \cite{508}. Besides, some authentification mechanisms need to be imposed, which, together with the highly directional energy beamforming, ensure that only the legitimated devices receive the significant wireless power.
\subsubsection {\it Inter-operate with wireless communication systems} radiative WPT systems need to have a minimal adversary impact on existing or future wireless communication systems. This can be achieved via two basic approaches. The first one is to develop standalone radiative WPT systems that are sufficiently isolated from existing communication systems in terms of spectrum usage, spatial separation, or interference mitigation, etc. For example, most prior radiative WPT designs use the 2.45GHz licence-free ISM (industrial, scientific, and medical)  band that has been originally reserved for various purposes other than telecommunications. On the other hand, wireless power and information transfer systems could be jointly designed to seamlessly integrate both, a  paradigm that has received tremendous research interests recently. There are mainly two lines of research under this paradigm, namely SWIPT (see \cite{745,746,747,534} and the references therein), where information and power are transmitted concurrently using the same RF waveform, and wireless powered communications (see \cite{515,525,744,753,742,743} and the references therein), where the energy for wireless communication at the devices is obtained via radiative WPT upon usage. The Wireless Identification and Sensing Platform (WISP) \cite{733} 
 and the Power over Wi-Fi systems \cite{734} both developed by University of Washington can be viewed as two practical implementations for low-power and low-duty-cycle wireless powered communication and SWIPT systems, respectively. However, more prototypes are needed to demonstrate the effectiveness of the SWIPT and wireless powered communication concepts for higher power applications (e.g., on the order of milliwatts and above).

\subsubsection {\it Safety and health guarantees} Radiative WPT systems can only be widely deployed if the safety and health issues are satisfactorily resolved. Compared to wireless communications, complying with the various authority regulations to ensure safety and health imposes more design challenges in WPT systems, owing to the higher transmission power needed in general.



\subsection{Objective and Organization}
The main objective of this article is to give a systematic treatment on the new communication and signal design techniques that can be applied for achieving efficient WPT. As can be seen from the preceding subsection, efficient wireless power and communication systems share several similar characteristics and hence the use of similar techniques, such as MIMO beamforming, closed-loop operation, transmitter coordination, etc. However, most of the existing techniques developed for wireless communications cannot be directly applied in WPT systems, due to their distinct design objectives (e.g., rate versus energy maximization), different practical limitations  (e.g., hardware and power constraints), as well as the different receiver sensitivities and models (linear versus non-linear). This article thus differs from the vast majority of the literature, which either treats WPT from the hardware design perspective, e.g., designing highly efficient rectennas \cite{749,748,754,755}, or considers the joint wireless power and information transmission where complicated compromise between the two 
 as well as some over-simplified assumptions on WPT have to be made. Instead, this article aims to address the various  specific requirements for WPT systems envisioned in the preceding subsection,  by leveraging the use of advanced communications and signals design techniques and exploiting the unique characteristics of WPT systems. 





\begin{figure}
\centering
\includegraphics[scale=0.5]{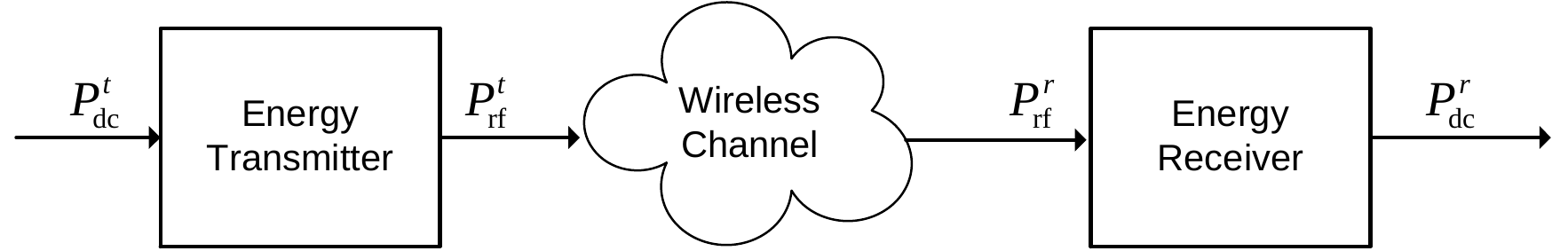}
\caption{The block diagram of a generic WPT system.}\label{F:BasicArchitecture}
\end{figure}


Fig.~\ref{F:BasicArchitecture} shows a generic WPT system, which consists of an energy transmitter (ET) and an energy receiver (ER) that are separated by a wireless medium.  At the ET, the DC (or low-frequency AC) energy-bearing signal is up-converted into the RF signal in a designated frequency band and radiated into the air by using transmitting antenna or antenna array. After propagating via the wireless channel, the RF signal arriving at the ER is picked up by the receiving antenna or antenna array, and then converted into usable DC power by rectifier. Denote by $P_{\dc}^t$ and $P_{\rf}^t$ the input DC power and output RF power at the ET, and $P_{\rf}^r$ and $P_{\dc}^r$ the input RF power and output DC power at the ER, respectively. The end-to-end power transfer efficiency $e$ can be  expressed as
\begin{align}
e=\frac{P_{\dc}^r}{P_{\dc}^t}=\underbrace{\frac{P_{\rf}^t}{P_{\dc}^t}}_{e_1}\underbrace{\frac{P_{\rf}^r}{P_{\rf}^t}}_{e_2}\underbrace{\frac{P_{\dc}^r}{P_{\rf}^r}}_{e_3},
\end{align}
where $e_1$, $e_2$, and $e_3$ denote the DC-to-RF, RF-to-RF, and RF-to-DC power conversion/transmission efficiency, respectively. 
 Under the assumption that the DC-to-RF conversion efficiency $e_1$ at the ET is fixed, this article will focus on the various communication and signal processing techniques for maximizing the DC output power $P_{\dc}^r$ at the ER. In this case,  both $e_2$ and $e_3$ need to be optimized, and they are in general coupled with each other due to the non-linearity of the energy rectification process at the ER. On the other hand, for scenarios with sufficiently weak incident RF power $P^r_{\rf}$, the rectification process can be approximated as linear, as will be seen in Section~\ref{sec:RectModel}, i.e., $e_3$ is fixed regardless of the input power and waveform. In this case, maximizing $P_{\dc}^r$ reduces to maximizing the incident RF power $P_{\rf}^r$, or equivalently the RF-to-RF transmission efficiency $e_2$.

The rest of this article is organized as follows. In Section~\ref{sec:RectModel}, we will present a simple and tractable model of the rectenna circuit and derive the generic output DC power at the ER after rectification. Under the assumption of linear rectification at the ER, Sections~\ref{sec:Pt2Pt} and \ref{sec:WPTNetwork} will focus on various techniques on improving the received RF power at the ER for single- and multi-user WPT systems, respectively. In Section~\ref{Section_NL}, the general non-linear energy harvesting model will be adopted, where the power waveforms are optimized by exploiting the receiver non-linearity.  Section~\ref{sec:FurtherDiscussion} extends discussions on various other issues pertaining to the design and implementation of WPT systems. Lastly, Section~\ref{sec:Conclusion} concludes the paper.

\emph{Notations:} In this paper, scalars are denoted by italic letters. Boldface lower- and upper-case letters denote vectors and matrices, respectively. $\mathbb{C}^{M\times N}$ denotes the space of $M\times N$ complex matrices. $\jImag$ denotes the imaginary unit, i.e., $\jImag^2=-1$. $\mathbb{E}[\cdot]$ denotes the statistical expectation and $\Re\{\cdot\}$ represents the real part of a complex number. $\mathbf{I}_M$ denotes an $M\times M$ identity matrix and $\mathbf{0}$ denotes an all-zero vector/matrix. For an arbitrary-size matrix $\mathbf{A}$,  its complex conjugate, transpose, Hermitian transpose, and Frobenius  norm are respectively denoted as $\mathbf A^*$, $\mathbf{A}^{T}$, $\mathbf{A}^{H}$ and $\|\mathbf{A}\|_F$. $[\mathbf A]_{im}$ denotes the $(i,m)$th element of matrix $\mathbf A$. For a square Hermitian matrix $\mathbf{S}$, $\mathrm{Tr}(\mathbf{S})$ denotes its trace, while $\lambda_{\max}(\mathbf S)$ and $\mathbf v_{\max}(\mathbf S)$ denote its largest eigenvalue and the corresponding eigenvector, respectively.

\section{Analytical Model of the Rectenna}\label{sec:RectModel}
A rectenna harvests ambient EM energy, then rectifies and filters it (using a diode and a low-pass filter). The recovered DC power then either powers a low-power device directly, or is stored in a super-capacitor or battery for higher power and low-duty-cycle operation.

\subsection{Antenna Model}
\begin{figure}
 \begin{minipage}[c]{.5\linewidth}
   \centerline{\includegraphics[width=\columnwidth]{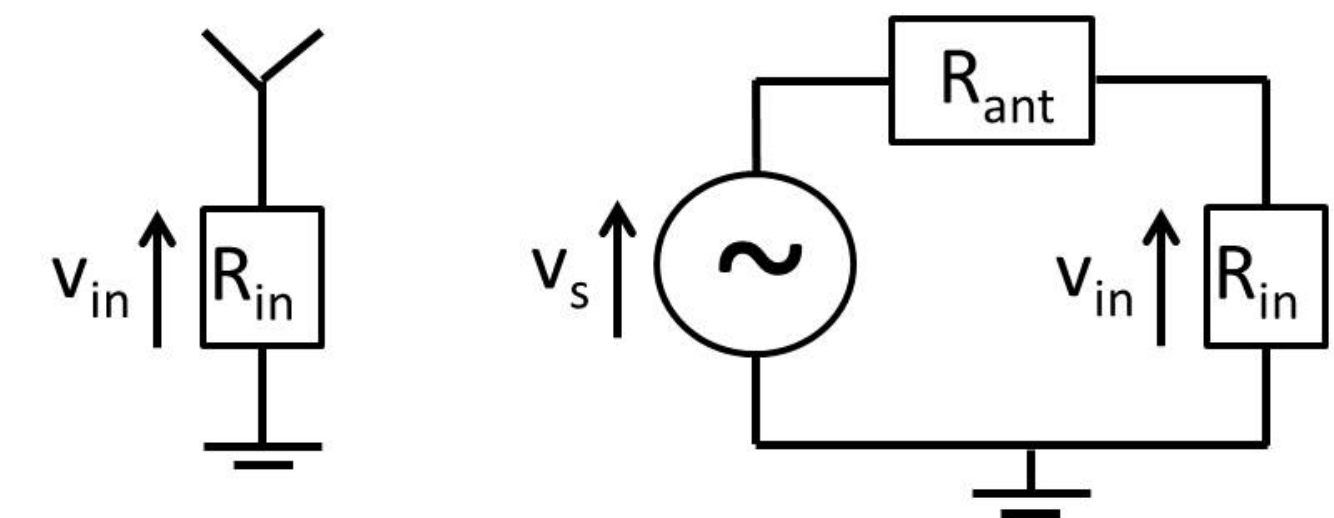}}
  \end{minipage}\hfill
 \begin{minipage}[c]{.5\linewidth}
   \centerline{\includegraphics[width=\columnwidth]{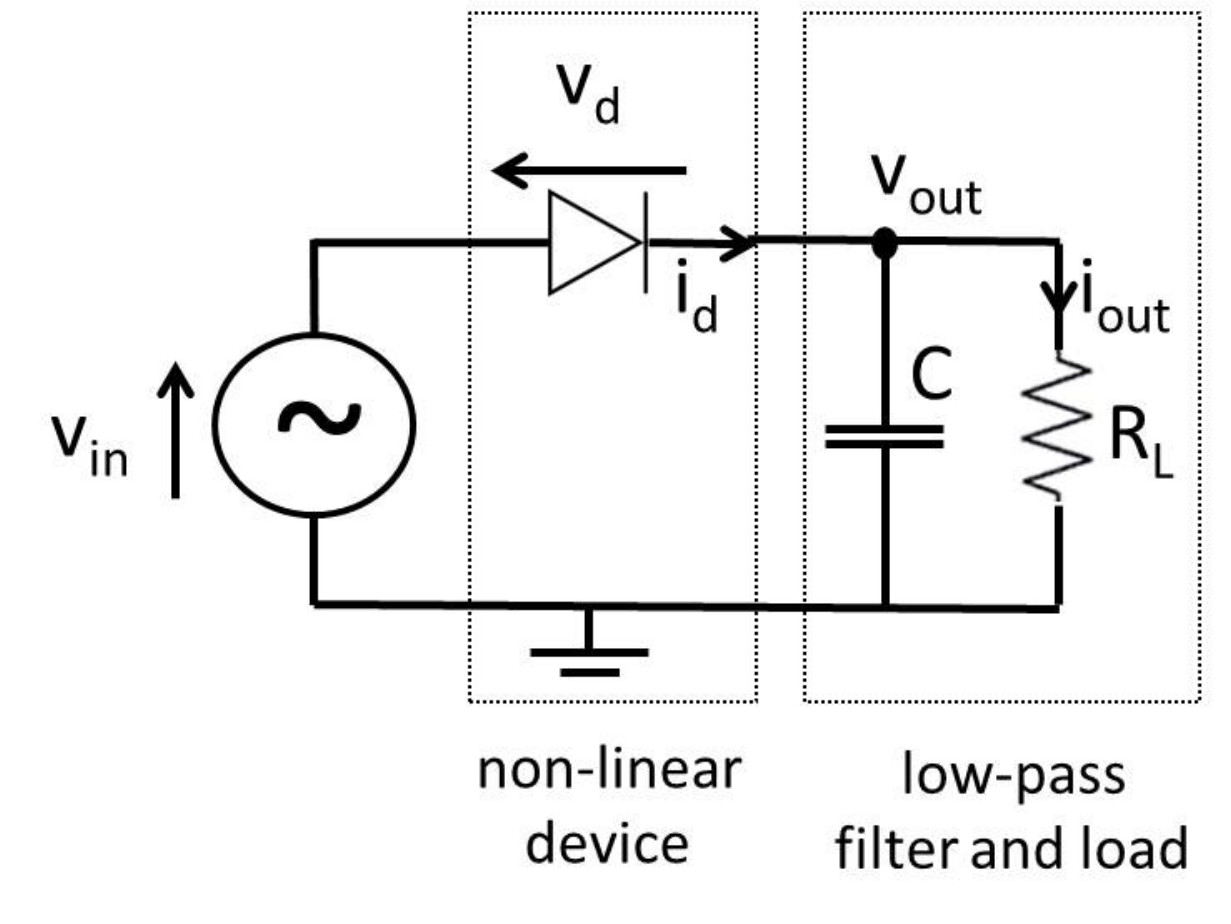}}
  \end{minipage}
  \caption{Antenna equivalent circuit (left) and a single diode rectifier (right).}
  \label{antenna_model}
\end{figure}

The antenna model reflects the power transfer from the
antenna to the rectifier through the matching network. As illustrated in Fig. \ref{antenna_model}(left), a lossless antenna
can be modelled as a voltage source $v_s(t)$ followed by a
series resistance $R_{ant}$. Let $Z_{in} = R_{in} + \jImag X_{in}$ denote the
input impedance of the rectifier with the matching network.
Assuming perfect matching ($R_{in} = R_{ant}$, $X_{in} = 0$), all the
available RF power $P_{\textnormal{rf}}^r$ is transferred to the rectifier and
absorbed by $R_{in}$, so that $P_{\textnormal{rf}}^r = \mathbb{E}\big[\left|v_{in}(t)\right|^2\big]/R_{in}$ and $v_{in}(t)=v_{s}(t)/2$. Since $P_{\textnormal{rf}}^r =\mathbb{E}\big[\left|y(t)\right|^2\big]$ with $y(t)$ denoting the RF signal impinging on the rectenna, $v_{in}(t)$ can be formed as
\begin{equation}\label{vs_eq}
v_{in}(t)=y(t)\sqrt{R_{in}}=y(t)\sqrt{R_{ant}}.
\end{equation}

\subsection{Rectifier and Diode Models}
Consider a single receive antenna ($M_r=1$) and a rectifier composed of a single series diode followed by a low-pass filter with load as in Fig. \ref{antenna_model}(right). Denoting the voltage drop across the diode as $v_d(t)=v_{in}(t)-v_{out}(t)$ where $v_{in}(t)$ is the input voltage to the diode and $v_{out}(t)$ is the output voltage across the load resistor, a tractable behavioural diode model is obtained by Taylor series expansion of the diode characteristic equation $i_d(t)=i_s \big(e^{\frac{v_d(t)}{n_f v_t}}-1 \big)$ (with $i_s$ the reverse bias saturation current, $v_t$ the thermal voltage, $n_f$ the ideality factor assumed equal to $1.05$) around a quiescent operating point $v_d=a$, namely
\begin{equation}\label{DiodeCurrent}
i_d(t)=\sum_{i=0}^{\infty }k_i' \left(v_d(t)-a\right)^i,
\end{equation}
where $k_0'=i_s\big(e^{\frac{a}{n_f v_t}}-1\big)$ and $k_i'=i_s\frac{e^{\frac{a}{n_f v_t}}}{i!\left(n_f v_t\right)^i}$, $i=1,\ldots,\infty$.

\par Assume a steady-state response and an ideal low-pass filter such that $v_{out}(t)$ is at constant DC level. Choosing $a=\mathbb{E} \left[ v_d(t) \right]=-v_{out}$ and using \eqref{vs_eq}, \eqref{DiodeCurrent} can be simplified as
\begin{equation}\label{polynomialSeries}
i_d(t)=\sum_{i=0}^{\infty }k_i' v_{in}(t)^i=\sum_{i=0}^{\infty }k_i' R_{ant}^{i/2} y(t)^i.
\end{equation}
Note that $a=-v_{out}$ highlights that the diode is negatively biased due to the output voltage $v_{out}$ across the load resistor being greater than zero.
The DC output power is directly proportional to the DC component of the current flowing through the load. The DC component of $i_{d}(t)$ is the time average of the diode current, and is obtained as
\begin{equation}\label{polynomialSeries_DC}
i_{out}=\sum_{i \hspace{0.1cm}\textnormal{even}}^{\infty }k_i' R_{ant}^{i/2} \mathbb{E}\left[y(t)^i\right].
\end{equation}
There are no odd-order terms since $\mathbb{E}\left[y(t)^i\right]=0$ for $i$ odd. More details on this model can be found in \cite{Clerckx:2016b}.

\par Throughout the paper, the aim from a system design perspective will be to find transmission strategies that maximize $i_{out}$ subject to a transmit RF power constraint. This may appear as a challenging problem since the rectifier characteristics ${k_i'}$ are functions of $a=-v_{out}=-R_L i_{out}$ in the Taylor expansion and therefore a function of the output DC current $i_{out}$. Making this dependence explicit, we can write $i_{out}$ in \eqref{polynomialSeries_DC} as
\begin{equation}\label{i_out_dependence}
i_{out}\approx \sum_{i \hspace{0.1cm}\textnormal{even}}^{n_o} k_i'\left(i_{out}\right) R_{ant}^{i/2} \mathbb{E}\left[y(t)^i\right]
\end{equation}
where we truncated the Taylor expansion to order $n_o$, and $n_o$ is an even integer with $n_o\geq 2$.
Fortunately, it is shown in \cite{Clerckx:2016b} that from a transmit signal/waveform optimization perspective, maximizing $i_{out}$ in \eqref{i_out_dependence} (subject to a transmit RF power constraint) is equivalent to maximizing the quantity
\begin{equation}\label{z_DC_def}
z_{DC}=\sum_{i \hspace{0.1cm}\textnormal{even}, i\geq 2}^{n_o} k_i R_{ant}^{i/2} \mathbb{E}\left[y(t)^i\right]
\end{equation}
where $k_i=\frac{i_s}{i!\left(n_f v_t\right)^i}$. Parameters $k_i$ and $z_{DC}$ are now independent of the quiescent operating point $a$.
Leveraging \eqref{z_DC_def}, we can now define two types of rectifier model.
\par Let us first truncate \eqref{z_DC_def} to order 2 ($n_o=2$) such that $z_{DC}= k_2 R_{ant} \mathbb{E}\left[y(t)^2\right]$. We note that $z_{DC}$ writes as a linear function of $\mathbb{E}\left[y(t)^2\right]$. This is the rectifier \textit{linear model}. Interestingly, finding the best transmit strategy so as to maximize $z_{DC}$, subject to a transmit RF power constraint, is equivalent to the one that maximizes $\mathbb{E}\left[y(t)^2\right]$ \cite{514}. Therefore, for a second-order truncation,  the model of the rectifier is linear, which gives a reasonable approximation for sufficiently low input RF power when the higher-order terms would not contribute relatively much to $z_{DC}$. In this case, maximizing $e_2\times e_3$ corresponds to maximizing $e_2$ with constant $e_3$, or equivalently the transmission strategy that maximizes the RF power at the input to the rectifier is the same strategy that maximizes the DC output current (and therefore DC output power). The linear energy harvesting model will be assumed in Sections~\ref{sec:Pt2Pt} and \ref{sec:WPTNetwork}.
\par Let us now truncate \eqref{z_DC_def} to a higher-order term, e.g.\ order 4 ($n_o=4$) for simplicity. This is a \textit{nonlinear model} of the rectifier. Quantity $z_{DC}$ is now approximated as
\begin{equation}\label{polynomialSeries_DC_4}
z_{DC}= k_2 R_{ant}\mathbb{E}\left[y(t)^2\right]+k_4 R_{ant}^2\mathbb{E}\left[y(t)^4\right].
\end{equation}
The non-linearity of the rectifier is now characterized through the presence of the fourth-order term $\mathbb{E}\left[y(t)^4\right]$. As it will appear clearer in Section \ref{Section_NL}, maximizing $z_{DC}$ or equivalently $e_2 \times e_3$ does not lead to the same solution as maximizing $e_2$ only.

\section{Single-User WPT}\label{sec:Pt2Pt}
In Sections~\ref{sec:Pt2Pt} and \ref{sec:WPTNetwork}, we will present the various techniques for efficient WPT under the linear energy harvesting model, i.e., with constant RF-to-DC power conversion efficiency $e_3$.
\subsection{System Model}\label{sec:SUModel}
We first consider a single-user point-to-point  MIMO WPT system in the general multi-path environment, where an ET equipped with $M_t\geq 1$ antennas transmits RF power wirelessly to an ER with $M_r\geq 1$ antennas. We consider the most general setup of multi-band WPT, with the commonly used single-band or single-tone power transmission as a special case. We assume that a total of $N$ orthogonal sub-bands are used, where the $n$th sub-band has carrier frequency $f_n$ and equal bandwidth $B_s$, $n=1,\cdots, N$. Therefore, the signal transmitted by antenna $m$ can be expressed as
\begin{align}
x_m(t) & = \sqrt{2} \sum_{n=1}^N a_{mn}(t)\cos \big(2\pi f_n t + \phi_{mn}(t)\big),\notag \\
& =\sqrt{2} \Re \left\{\sum_{n=1}^N s_{mn}(t) e^{\jImag 2\pi f_n t} \right \},\  m=1,\cdots, M_t,\label{eq:xm}
\end{align}
where $s_{mn}(t)\triangleq a_{mn}(t)e^{\jImag \phi_{mn}(t)}$ with signal bandwidth no greater than $B_s$ denotes the complex-valued baseband signal transmitted by antenna $m$ at sub-band $n$. For the special case of unmodulated WPT,  $s_{mn}(t)$ is constant across $t$, i.e., $s_{mn}(t)=s_{mn}=a_{mn}e^{\jImag \phi_{mn}}$, $\forall t$. In this case, $x_m(t)$ is a summation of $N$ sinewaves inter-separated by $B_s$ Hz, and hence essentially occupies zero bandwidth.

Let $L$ denote the number of multipaths between the ET and ER, $\alpha_l$ and $\tau_l$ be the amplitude gain and delay of the $l$th path, respectively. 
Further denote by $\xi_{imnl}$ the phase shift of the $l$th path between transmit antenna $m$ and receive antenna $i$ at subcarrier $n$, whose value depends on the array configuration, the angle of departure/arrival (AoD/AoA) of the $l$th path, as well as the carrier frequency $f_n$. The signal received at antenna $i$ due to transmit antenna $m$ can then be expressed as
\begin{align}
y_{im}(t)&=\sqrt{2} \Re \left\{\sum_{l=1}^L \sum_{n=1}^N  \alpha_l s_{mn}(t-\tau_l) e^{\jImag \xi_{imnl}} e^{\jImag 2\pi f_n (t-\tau_l)} \right\} \notag \\
&\approx \sqrt{2} \Re \left\{\sum_{n=1}^N h_{imn}^* s_{mn}(t)  e^{\jImag 2\pi f_nt} \right\},
\end{align}
where we have assumed $\underset{l\neq l'}{\max}|\tau_l-\tau_{l'}|\ll 1/B_s$ so that $s_{mn}(t)$ for each sub-band $n$ is a narrowband signal, thus $s_{mn}(t-\tau_l)\approx s_{mn}(t)$, $\forall l$, and
$h_{imn}^* \triangleq \sum_{l=1}^L \alpha_l e^{\jImag \xi_{imnl}}e^{-\jImag 2\pi f_n\tau_l}$
denotes the flat-fading channel between transmit antenna $m$ and receive antenna $i$ at sub-band $n$. The total received signal at antenna $i$ is a superposition of those from all the $M_t$ transmit antennas, i.e.,
\begin{align}
y_i(t) & =  \sum_{m=1}^{M_t} y_{im}(t)\notag \\
&=  \sqrt{2}  \Re \left\{  \sum_{n=1}^N \mathbf h_{in}^H \mathbf s_n(t) e^{\jImag 2\pi f_nt}  \right\}, \ i=1,\cdots, M_r,\label{eq:yit}
\end{align}
where $\mathbf h_{in}^H \triangleq \left[h_{i1n}^*, \cdots, h_{iM_tn}^*\right]$ denotes the channel vector from the $M_t$ transmit antennas to receive antenna $i$ at sub-band $n$, and $\mathbf s_n(t) \triangleq \left[s_{1n}(t), \cdots  s_{M_tn}(t)\right]^T$ denotes the  signals  transmitted  by the $M_t$  antennas at sub-band $n$.
The total RF power received by all the $M_r$ antennas of the ER can then be expressed as
\begin{align}
P_{\rf}^r&=\sum_{i=1}^{M_r} \mathbb{E} \left[y_i(t)^2\right]=\sum_{i=1}^{M_r} \sum_{n=1}^N \mathbb{E} \left[|\mathbf h_{in}^H \mathbf s_n(t)|^2 \right]\notag \\
&=\sum_{n=1}^N \tr\left(\mathbf H_n^H \mathbf H_n \mathbf S_n\right),\label{eq:prfr}
\end{align}
where $\mathbf H_n^H \triangleq \left [\mathbf h_{1n},\cdots, \mathbf h_{M_rn}\right]\in \mathbb{C}^{M_t\times M_r}$ denotes the MIMO channel matrix from the $M_t$ transmit antennas to the $M_r$ receive antennas at sub-band $n$, and $\mathbf S_n \triangleq \mathbb {E} \left[\mathbf s_n(t) \mathbf s_n^H(t) \right]\in \mathbb{C}^{M_t\times M_t}$ is a positive semidefinite matrix denoting the transmit covariance matrix at sub-band $n$. Without loss of generality for WPT, we assume that $\mathbf s_n(t)$ constitutes pseudo-random signals.\footnote{If $\mathbf s_n(t)$ is used for the dual purposes of both wireless power and information transmissions as in the SWIPT setup, it needs to be designed by taking into account the practical modulation scheme used in wireless communications.}  Note that for the special case of unmodulated WPT with $s_{mn}(t)$ being deterministic, we have $\mathbf S_n=\mathbf s_n \mathbf s_n^H$, which is constrained to be a rank-1 matrix. Thus, as compared to unmodulated transmission, modulated WPT offers more design freedom  by enabling multi-beam transmission since $\mathbf S_n$ could be of arbitrary rank no greater than $M_t$.



  The RF power transmitted by the ET is 
 \begin{align}
 P_{\rf}^t=\sum_{m=1}^{M_t} \mathbb{E}[x_m(t)^2]= \sum_{n=1}^N \tr(\mathbf S_n),\label{eq:prft}
 \end{align}
 with $\tr(\mathbf S_n)$ being the transmit power at sub-band $n$.


Under the linear energy harvesting model, the  RF-to-DC energy conversion efficiency $e_3$  is a constant. As a result, the amount of DC power harvested by the ER is then simply given by $P_{\dc}^r=e_3P_{\rf}^r$. In this case, maximizing $P_{\dc}^r$ is equivalent to maximizing the received RF power $P_{\rf}^r$ via optimizing the transmit covariance matrices $\mathbf S_n$ over the $N$ sub-bands.


\subsection{Energy Beamforming}\label{sec:EnergyBF}
The power maximization problem based on \eqref{eq:prfr}  and \eqref{eq:prft} can be formulated as
\begin{equation}\label{P:SingleUser}
\begin{aligned}
 \quad \  \underset{\{\mathbf S_n\}}{\max} \ & \sum_{n=1}^N \tr \left( \mathbf H_n^H \mathbf H_n \mathbf S_n\right) \\
\text{s.t.}\ & \sum_{n=1}^N \tr \left(\mathbf S_n\right) \leq P_{\rf}^t, \\
& \tr \left(\mathbf S_n\right) \leq P_s, \ \forall n,\\
& \mathbf S_n \succeq \mathbf 0, \ \forall n,
\end{aligned}
\end{equation}
where $P_{\rf}^t$ denotes the total transmit power constraint at the ET across all the $N$ sub-bands, and $P_s$ is the transmit power limit at each frequency sub-band, which could correspond to the power spectrum density constraint imposed by the regulatory authorities \cite{549}. For instance, according to the FCC (Federal Communications Commission) regulations Part 15.247, paragraph (e): the power spectrum density over the 902-928MHz band from the intentional radiator ``shall not be greater than 8dBm in any 3kHz band'' \cite{549}. Thus, the per-sub-band power limit $P_s$ not only depends on the bandwidth $B_s$, but also on how the power is distributed across the spectrum. In particular, compared to unmodulated WPT where the signal power is concentrated on discrete frequency tones, modulated WPT usually has more relaxed $P_s$ since the signal power of each sub-band is spread across the spectrum of bandwidth $B_s$. Therefore, modulated WPT is in general preferable for high-power delivery.  Without loss of generality, we assume that $P_s\leq P_{\rf}^t \leq N P_s$, since otherwise, either the sum-power constraint or the per-sub-band power constraint in \eqref{P:SingleUser} is redundant and hence can be removed. In addition, for the convenience of exposition, we assume that $P_{\rf}^t$ is an integer multiple of $P_s$, i.e., $P_{\rf}^t/P_s=N'$ for some integer $1\leq N'\leq N$.

For any given power allocation $p_n=\tr(\mathbf S_n)$, it is not difficult to verify that the optimal covariance matrix $\mathbf S_n$ to \eqref{P:SingleUser} should be
\begin{align}
\mathbf S_n = p_n \mathbf v_n \mathbf v_n^H, \ n=1,\cdots, N,\label{eq:Sn}
\end{align}
where $\mathbf v_n= \mathbf v_{\max}(\mathbf H_n^H \mathbf H_n)$ denotes the eigenvector corresponding to the dominant eigenvalue of $\mathbf H_n^H \mathbf H_n$. The resulting received power at each sub-band $n$ is
\begin{align}
P_{\rf, n}^r= p_n \lambda_{\max,n}, \ n=1,\cdots, N,
\end{align}
where $\lambda_{\max,n}= \lambda_{\max}(\mathbf H_n^H \mathbf H_n)$ denotes the maximum eigenvalue of $\mathbf H_n^H \mathbf H_n$ for sub-band $n$. As a result, problem \eqref{P:SingleUser} reduces to
\begin{equation}\label{P:SingleUser2}
\begin{aligned}
 \quad \  \underset{\{p_n\}}{\max} \ & \sum_{n=1}^N p_n \lambda_{\max,n} \\
\text{s.t.}\ & \sum_{n=1}^N p_n \leq P_{\rf}^t, \\
& p_n \leq P_s, \ \forall n,\\
& p_n \geq 0, \ \forall n.
\end{aligned}
\end{equation}

Problem \eqref{P:SingleUser2} is a simple linear programming (LP), whose optimal solution is given by
\begin{align}\label{eq:pn}
p_{[n]}= \begin{cases}
P_s, \ & n=1,\cdots, N', \\
0, \ & n=N'+1, \cdots, N,
\end{cases}
\end{align}
where $[\cdot]$ is the permutation over all the $N$ sub-bands such that $\lambda_{\max,[1]}\geq \lambda_{\max,[2]}\cdots \geq \lambda_{\max,[N]}$. The corresponding optimal value of problem \eqref{P:SingleUser} is thus given by
\begin{align}\label{eq:Qmax}
P_{\rf}^{r}=P_s \sum_{n=1}^{N'} \lambda_{\max,[n]} .
\end{align}

It is observed from \eqref{eq:pn} that for MIMO multi-band WPT systems over frequency-selective channels under linear energy harvesting model, the optimal scheme is to transmit over the  $N'\leq N$ strongest sub-bands only, each with the maximum allowable power $P_s$. As a result, the remaining $N-N'$ unused sub-bands  could be opportunistically re-used for other applications such as information transmission.  The solution in \eqref{eq:Sn} also shows  that for each of the $N'$ strongest sub-bands, $\mathbf S_n$ is a rank-1 covariance matrix, 
i.e., unmodulated signal with single-beam transmission is optimal at each sub-band. In this case, the energy signals are only beamed towards the strongest eigenmode of the corresponding MIMO channel $\mathbf H_n$, regardless of the transmission power level. This is in  sharp contrast to conventional multi-band MIMO wireless communications, where in general all the spatial eigenmodes need to be utilized to fully realize the multiplexing gain if the transmit power is sufficiently large \cite{56}.  The expression in \eqref{eq:Qmax} shows that for multi-antenna  WPT systems in frequency-selective channels, both {\it frequency-diversity} as well as {\it energy beamforming} gains can be achieved to maximize the power transfer efficiency.

Note that if $P_{\rf}^t=P_s$ or $N'=1$, only the single strongest sub-band is used for power transfer, and the result in \eqref{eq:Qmax} can be more explicitly expressed as
\begin{align}\label{eq:Qmax2}
P_{\rf}^{r}=P_{\rf}^t \underset{n=1,\cdots N}{\max} \lambda_{\max}(\mathbf H_n^H \mathbf H_n).
\end{align}
This is different from the case of non-linear energy harvesting model as will be studied in Section~\ref{Section_NL}, where the power is in general allocated over more than one frequency sub-channels, not only on the one with the largest dominant eigenvalue.




\begin{figure}
\centering
\includegraphics[scale=0.45]{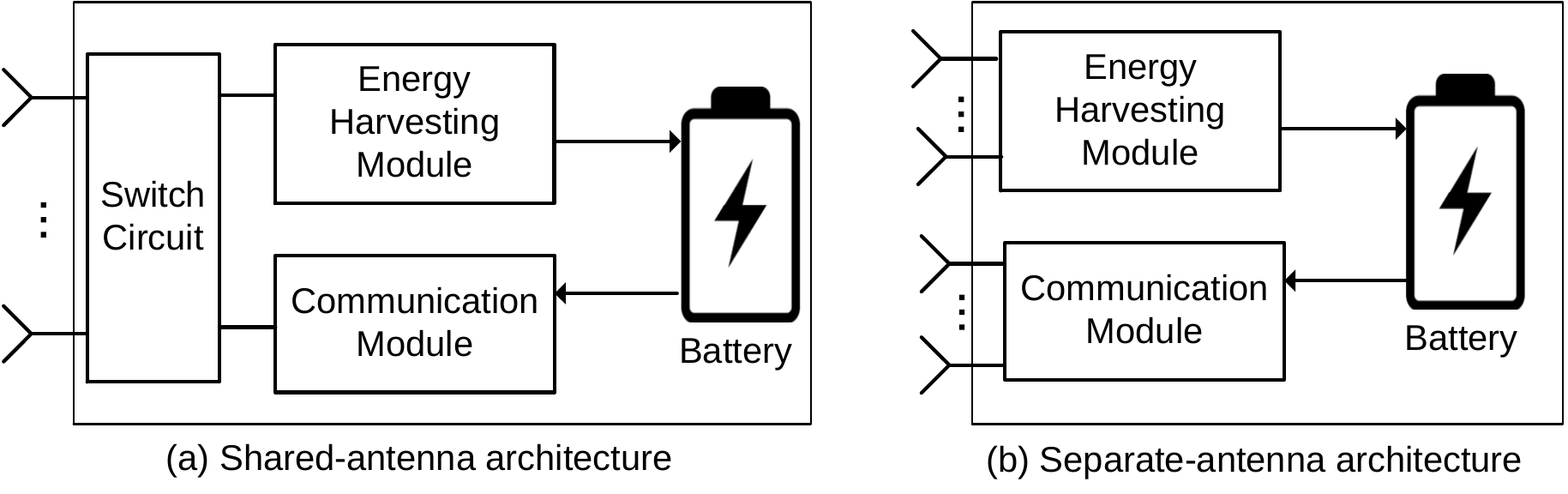}
\caption{Energy receiver with shared- versus separate-antenna architecture for energy harvesting and communication.}\label{F:SharedandSeparatedAnts}
\end{figure}

\subsection{Channel Acquisition}\label{sec:channelEstimation}
Both frequency-diversity and energy-beamforming gains shown in the preceding subsection critically depend on the channel state information (CSI) at the ET (CSIT). In principle, CSIT in WPT systems could be acquired with similar techniques as those developed in wireless communication systems \cite{484}. However,  WPT systems possess some unique characteristics, which need to be taken into account for designing efficient channel acquisition schemes tailored for power transmission, as discussed in the following.

{\it CSI at receiver:} In contrast to  communication systems, which usually require CSI to be also available at the receiver for coherent signal demodulation/detection, receiver-side CSI is in general unnecessary for WPT systems, since the arriving RF signal at the ER is directly converted to the DC power by rectifiers without requiring any signal processing to be applied.

{\it Net harvested energy:} Due to the energy scarcity at the ER, an efficient channel acquisition scheme for WPT systems needs to take into account the ER's energy consumption due to channel training and feedback. To achieve an optimal tradeoff between beamforming gain and the associated energy overhead, a useful design objective could be maximizing the {\it net harvested energy}, which is defined as the amount of harvested energy at the ER offset by that consumed for CSI acquisition \cite{528}.

{\it Hardware constraint:}  The CSI acquisition design for WPT systems may also need to take into account the limited hardware processing capability of ER. For example, for WPT in wireless sensing applications, the low-cost ERs in the sensors may not have the sophisticated channel estimation or signal processing capabilities as in conventional wireless communication systems, which calls for more innovative channel acquisition methods for WPT.

To facilitate the introduction of  the various channel acquisition schemes for WPT systems, we first classify the ER architectures based on whether the energy harvesting and the communication modules share the same set of antennas \cite{777}. Note that depending on the functionalities of the ER nodes, the communication modules could be either the built-in components of the ERs, or the dedicated modules specifically designed for enhancing the WPT performance via closed-loop operations. As shown in Fig.~\ref{F:SharedandSeparatedAnts}(a), with the {\it shared-antenna} architecture, the same set of antenna elements are connected to both the energy harvesting and communication modules via RF switches; thus, energy harvesting and communication take place in a time-division manner using the same antennas. Such an architecture has the merits of a more compact receiver form factor, easier channel estimation, etc. On the other hand, for the {\it separate-antenna} architecture as shown in Fig.~\ref{F:SharedandSeparatedAnts}(b), the energy harvesting and communication modules use distinct antennas, and thus they could be operated concurrently and independently. In the following, we first present the forward-link and reverse-link training based channel estimation schemes for the shared-antenna ER architecture, and then the power-probing scheme with limited energy feedback for the separate-antenna architecture. For simplicity, we consider narrow-band WPT ($N=1$) in the rest of this subsection.

\begin{figure}
\centering
\includegraphics[scale=0.6]{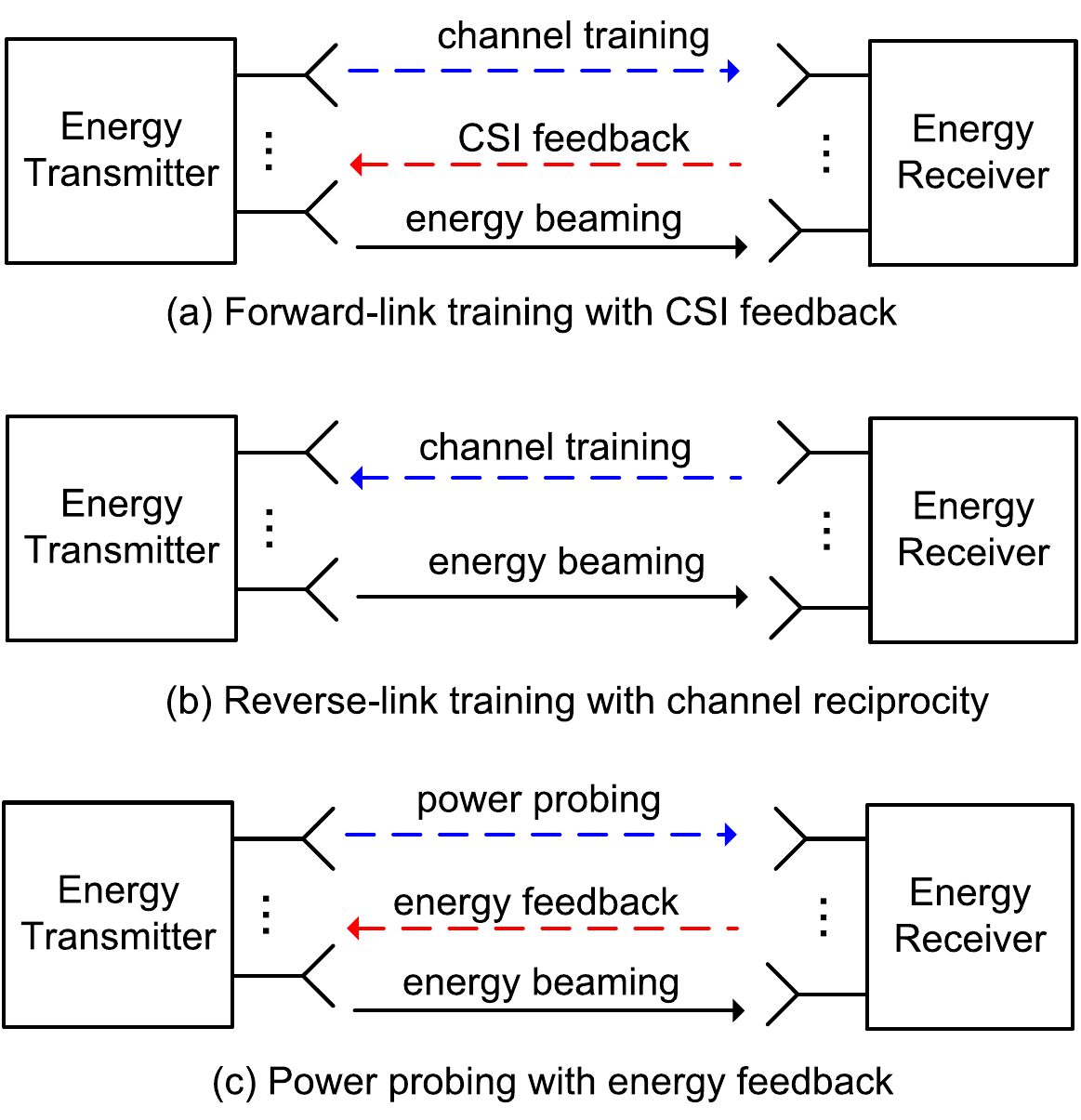}
\caption{Three channel acquisition schemes for WPT.}\label{F:ThreeCSILearningSchemes}
\end{figure}

\subsubsection{Forward-Link Training with CSI Feedback}
Similar to wireless communication systems, one straightforward approach to obtain CSI at the ET is by forward-link (from ET to ER) training together with reverse-link (from ER to ET) CSI feedback \cite{495,500,871,870}, as illustrated in Fig.~\ref{F:ThreeCSILearningSchemes}(a). With this scheme, pilot signals are sent from the ET to the ER, based on which the ER estimates the channel and sends the estimation back to the ET via a feedback link. Note that the CSI feedback could use different frequency from that for forward link training/energy transmission. However, in order to ensure that the estimated channel is indeed that used in subsequent energy transmission phase, such a scheme is only applicable for the shared-antenna ER architecture in Fig.~\ref{F:SharedandSeparatedAnts}(a). More importantly, its required training time increases with the number of antennas $M_t$ at the ET, and hence this method is not suitable when $M_t$ becomes large, such as for massive MIMO WPT systems. Besides, channel estimation at the ER requires complex baseband signal processing in general, which may not always be available at the ER for low-complexity nodes.



\subsubsection{Reverse-Link Training via Channel Reciprocity}\label{sec:reverseTraining}
An alternative channel acquisition method for shared-antenna WPT systems is via reverse-link training by exploiting the channel reciprocity \cite{528}, i.e., the channel matrices in the forward and reverse links between the ET and ER are assumed to be transpose of each other. Under this assumption, a fraction of the channel coherence time is assigned to the ER for sending pilot signals to the ET for direct channel estimation, as shown in Fig.~\ref{F:ThreeCSILearningSchemes}(b). Note that since ER itself does not require CSI for energy harvesting, no CSI feedback from the ET to ER is needed in general. Compared to the forward-link training-based scheme discussed previously, the reverse-link training-based scheme has two main advantages: (i) it is more efficient for large or massive MIMO WPT systems as the training overhead is independent of the number of antennas $M_t$ at the ET; (ii) it simplifies the processing at the ER since channel estimation and feedback operations are no longer required. However, this scheme greatly relies on the channel reciprocity assumption, which in practice requires accurate transmitter and receiver calibrations. It is worth noting that in wireless communication systems, reverse-link based training by assuming channel reciprocity is one of the key techniques for realizing massive MIMO systems to reduce the channel-acquisition overhead \cite{373}. However, the optimal reverse training design for WPT systems requires resolving the following new trade-off: too little training leads to coarsely estimated channel at the ET and hence reduced energy beamforming gain; whereas too much training consumes excessive energy harvested by the ER, and also leaves less time for energy transmission given a finite channel coherence time, thus resulting in less net harvested energy at the ER.

As a concrete example, we consider a MIMO point-to-point WPT system in narrow-band channel with $M_t$ antennas at the ET and $M_r$ antennas at the ER. For the purpose of exposition, we assume the simple quasi-static Rayleigh fading channel, for which the entries of the MIMO channel matrix $\mathbf H\in \mathbb{C}^{M_r \times M_t}$ are independent and identically distributed (i.i.d.) zero-mean circularly symmetric complex Gaussian (CSCG) random variables with variance $\beta$, i.e., $[\mathbf H]_{im}\sim \mathcal {CN}(0,\beta)$, $\forall i,m$. Note that the more general Rician fading channel systems are studied in \cite{528}. We further denote by $T$ the channel coherence block, i.e., the channel $\mathbf H$ is assumed to remain constant with the block of duration $T$, and varies independently from one block to another. As shown in Fig.~\ref{F:ThreeCSILearningSchemes}(b), each channel coherence block is divided into two phases: the reverse-link training phase with duration $\tau\leq T$, and the forward-link power transmission phase with duration $T-\tau$, for which the ET beams the wireless power to the ER based on the estimated channel. Without loss of generality, denote by $M_r'\leq M_r$ the number of antennas at the ER that participate in channel training, since not all the ER antennas should be trained if $M_r$ is large whereas $T$ is small. 
 Further denote by $p_{r}$ the training power sent by the ER during the reverse-link training phase. The total energy consumption at the ER for channel training is thus given by $p_r\tau$. On the other hand, it has been derived in \cite{528} that the average harvested energy at the ER (by assuming $e_3=1$ for notational convenience) with the above training-based scheme can be expressed as
\begin{align}
\bar Q(M_r',\tau,p_r)  = (T-\tau)P_{\rf}^t \beta & \Big(\frac{ p_r \tau \beta \Lambda(M_t, M_r')+\sigma^2 M_{r}'^2}{p_r\tau\beta+\sigma^2M_{r}'}\notag \\
&+M_r-M_r'\Big),\label{eq:QbarFinalRayleigh}
\end{align}
where $P_{\rf}^t$ is the transmission power by the ET during the energy transmission phase, $\sigma^2$ is the noise power at the ET during reverse-link training phase, and $\Lambda(M_t, M_r')\triangleq \mathbb{E}_{\mathbf X}\left[\lambda_{\max}(\mathbf X^H \mathbf X)\right]$, with $\mathbf X\in \mathbb{C}^{M_r'\times M_t}$ denoting the random matrix with i.i.d. zero-mean unit-norm CSCG entries, i.e., $[\mathbf X]_{im}\sim \mathcal{CN}(0,1)$, $\forall i,m$. Note that $\Lambda(M_t, M_r')$ monotonically increases with $M_t$ and $M_r'$. In the special cases of $M_t=1$ or $M_r'=1$,  it can be easily obtained that
 $\Lambda\left(M_t,1\right)=M_t$ and $\Lambda\left(1,M_r'\right)=M_r'$. For general $M_t$ and $M_r'$, no closed-form expression for $\Lambda\left(M_t,M_r'\right)$ is available, whereas its numerical values can be easily computed, e.g., based on the algorithm proposed in \cite{476}.


 The average harvested energy $\bar Q$ in \eqref{eq:QbarFinalRayleigh} can be viewed as a summation of two terms. The first term, which monotonically increases  with the training energy $p_r\tau$ and the number of ET antennas $M_t$, is attributed to the  $M_{r}'$ trained ER antennas whose corresponding channel matrix is estimated at the ET. The second term is attributed to the $(M_r-M_r')$ un-trained ER antennas, which is independent of the number of ET antennas $M_t$ since no beamforming gain can be achieved for energy transmission over the associated channel.

 The net average harvested energy at the ER can then be written as
\begin{equation}\label{eq:QnetRayleigh}
\begin{aligned}
\bar Q_{\text{net}}(M_r',\tau,p_r)  = & \bar Q(M_r',\tau,p_r)-p_r\tau.
\end{aligned}
\end{equation}

%

The optimal training power $p_r$, training duration $\tau$, as well as the number of training antennas $M_r'$ at the ER can then be obtained for net energy maximization based on \eqref{eq:QnetRayleigh} (see the details in \cite{528}).


\subsubsection{Power Probing with Limited Energy Feedback}\label{sec:EnergyFeedback}
For ERs with separate-antenna architecture shown in Fig.~\ref{F:SharedandSeparatedAnts}(b), the above two pilot training based channel estimation schemes are no longer applicable. This is because with distinct antennas used for energy harvesting and communication modules, the channels corresponding to the antennas used for energy harvesting cannot be trained directly with the communication antennas at the ER. To resolve this issue, \cite{491} and \cite{776} proposed a novel channel learning method with limited feedback based on the harvested energy levels at the ER.

Fig.~\ref{F:ThreeCSILearningSchemes}(c) shows the basic process of MIMO point-to-point WPT based on limited energy feedback. It is assumed that the ER is equipped with an energy meter, which is able to accurately measure the amount of energy harvested by the ER for a certain time duration. Upon receiving energy request from the ER, the ET starts transmitting energy using a sequence of carefully designed transmit covariance matrices $\mathbf S_1$, $\cdots$, $\mathbf S_\tau$, with $\tau$ denoting the number of training intervals for the channel learning phase, each assumed to have length $T_s$ seconds. Thus, the harvested energy by the ER in the $i$-th training interval is given by
\begin{align}
Q_i=T_s \tr(\mathbf G \mathbf S_i), \ i=1,\cdots, \tau, \label{eq:Qi}
\end{align}
where $\mathbf G\triangleq \mathbf H^H \mathbf H\in \mathbb{C}^{M_t\times M_t}$ denotes the matrix to be learned at the ET.  At the end of each training interval $i$, the ER sends a feedback information $f_i$ of $B$ bits to the ET based on its present and past energy measurements $Q_1,\cdots, Q_i$. In other words, $f_i$ specifies the energy feedback scheme by the ER that is in general a mapping from $Q_1,\cdots, Q_i$ to a $B$-bits feedback signal.  Based on the received feedback $\{f_i\}_{i=1}^\tau$ and the transmit covariance matrices $\{\mathbf S_i\}_{i=1}^\tau$ applied during the channel learning phase, the ET can obtain an estimate of the MIMO channel $\mathbf G$. The key is then to jointly design the specific feedback scheme $\{f_i\}$ at the ER, as well as the probing covariance matrices $\{\mathbf S_i\}$ and the channel estimation scheme at the ET.

To illustrate this, we adopt the analytical center cutting plane method (ACCPM) \cite{200} with the simple one-bit feedback scheme ($B=1$) proposed in \cite{491} in the  following. A more general energy feedback design with $B>1$ based on energy level quantization and/or comparison can be found in \cite{776}. With $B=1$, the feedback information $f_i$ at the $i$th training interval is set by comparing the harvested energy level $Q_i$ with $Q_{i-1}$ as
\begin{align}
f_i=\begin{cases}
1, \ &\text{if } Q_{i}\leq Q_{i-1}\\
-1, \ &\text{otherwise}.
\end{cases}
\end{align}

It then follows from \eqref{eq:Qi} that the ET obtains the following equality upon receiving the feedback bit $f_i$:
\begin{align}
f_i \tr\left(\mathbf G(\mathbf S_i-\mathbf S_{i-1}) \right)\leq 0,
\end{align}
which can be regarded as a {\it cutting plane} of $\mathbf G$, i.e., $\mathbf G$ must lie in the half space of $\mathcal H_i=\{\mathbf G| f_i  \tr\left(\mathbf G(\mathbf S_i-\mathbf S_{i-1}) \right)\leq 0\}$. By denoting $\mathcal P_i$ the set that is known to contain the channel matrix $\mathbf G$ after training interval $i$, we then have $\mathcal P_i=\mathcal P_{i-1} \cap \mathcal H_i$, $\forall i\geq 2$, or equivalently
\begin{align}
\mathcal P_i = \left\{ \mathbf G | f_l \tr\left(\mathbf G(\mathbf S_l-\mathbf S_{l-1}) \right)\leq 0, 2\leq l\leq i
\right\}, \ i=2,\cdots, \tau.\label{eq:Pi}
\end{align}
It is evident that $\mathbf G \in \mathcal P_\tau \subseteq \mathcal P_{\tau-1} \cdots \subseteq \mathcal P_2$. Note that $\mathcal P_i$ in \eqref{eq:Pi} defines a sequence of polyhedrons with decreasing volume and all containing $\mathbf G$. The analytic center of $\mathcal P_i$, denoted as $\hat {\mathbf G}_i$, can be efficiently obtained by solving a convex optimization problem \cite{202}. With $\hat {\mathbf G}_i$ obtained at the ET, the probing transmit covariance matrix $\mathbf S_{i+1}$ at next training interval is then designed to ensure that the resulting cutting plane is at least {\it neutral}, i.e.,
$\tr \left(\hat{\mathbf G}_i \left( \mathbf S_{i+1}-\mathbf S_i\right)\right)=0$.
It is shown in \cite{491} that the above ACCPM based channel learning algorithm with simple one-bit energy feedback converges to the true channel matrix $\mathbf G$ with increasing $\tau$.

%








\subsection{Extension and Future Work}
\subsubsection{Retrodirective-Amplification WPT}\label{sec:retrodirective}
A low-complexity energy beamforming scheme without requiring explicit channel estimation/feedback is {\it retrodirective amplification}. Retrodirective  transmission is a simple beamforming technique for multi-antenna arrays, which, upon receiving a signal from any direction,  transmit a signal response back to the same direction without the need of knowing the source direction \cite{805}, \cite{806}. The main idea is to exploit channel reciprocity and transmit a phase-conjugated version of the received signal. This can be automatically achieved by {\it retrodirective arrays} without relying on sophisticated digital signal processing. Two well known retrodirective array structures are Van Atta arrays \cite{805} and the heterodyne retrodirective arrays with phase-conjugating circuits \cite{804}. For WPT systems, the same retrodirective principle can be applied to achieve low-complexity energy beamforming as well as coordinated multipoint (CoMP) energy transmission with distributed antennas. WPT using retrodirective techniques have been experimentally demonstrated in different setups \cite{807,808,809,810,811,812}. In practice, since the ET amplifies the received signal as well as the background noise, the retrodirective WPT needs to be designed to be robust to the noise effect. In particular, similar to the reverse-link channel estimation based WPT, the training power by the ER in the reverse link needs to be optimized to balance between the retrodirective energy beamforming gain and the energy consumption of the ER.

\subsubsection{Channel Acquisition in Frequency-Selective Channel}
Channel acquisition for WPT in multi-antenna frequency-selective channels is in general more challenging than its frequency-flat counterpart, since in this case, the channels both in space and frequency domains need to be estimated to reap the benefits of both energy beamforming and frequency-diversity gains, as given by \eqref{eq:Qmax}. In \cite{558}, a reverse-link training based channel estimation scheme is proposed for MISO multi-band frequency-selective WPT systems, where the training design is optimized to maximize the net harvested energy at the ER. However, the optimal training design for the general MIMO wide-band WPT systems remains an open problem. Besides, existing studies are mostly based on the assumption of independent channels in both spatial and frequency domains. For some practical setup with correlated channels,  the training design could exploit the spatial and/or frequency channel correlations to further reduce the training overhead and hence enhance the overall energy transfer efficiency, which needs further investigation. Besides, for ERs with separate-antenna architecture for energy harvesting and communication, the extension of the channel acquisition scheme with limited energy feedback to frequency-selective channels also requires further studies.




%
%
%
%
%
%
%
%
%
%

\section{Multi-User WPT}\label{sec:WPTNetwork}
\begin{figure}
\centering
\includegraphics[scale=0.45]{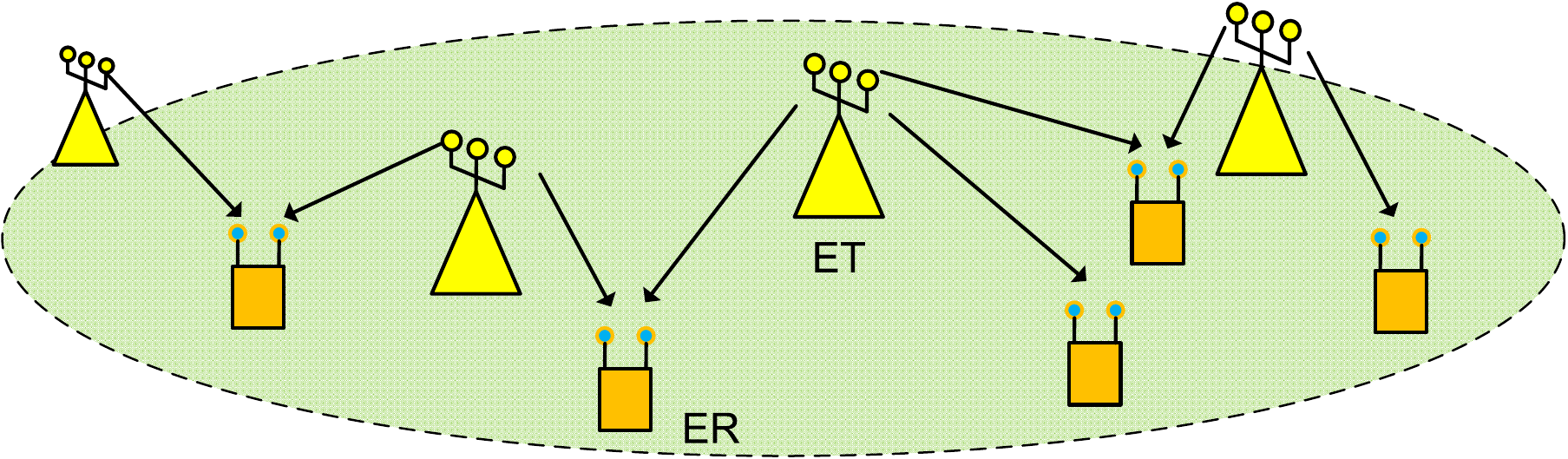}
\caption{Multi-user MIMO WPT system.}\label{F:MultiUserWPT}
\end{figure}

In practice, WPT systems generally need to simultaneously serve  $K\geq 1$ ERs with $J\geq 1$ distributed ETs in a network, as shown in Fig.~\ref{F:MultiUserWPT}. In this section, we consider a multi-user MIMO WPT system where each ET is equipped with $M_t\geq 1$ antennas and each ER with $M_r\geq 1$  antennas. 

\subsection{WPT Network Architecture}\label{sec:WPTArchiecture}
Similar to wireless communication networks, WPT systems  could have various networking architectures depending on the different levels of cooperation among the ETs.
\subsubsection{CoMP-based WPT} With CoMP-based WPT, all the $J$ ETs jointly design their energy signals to the $K$ ERs based on the global CSI of all the WPT links. This could be achieved by interconnecting  the ETs via high-capacity low-latency backhaul links to a central unit (CU), which is responsible for collecting the CSI from all ETs, optimizing the transmit signals based on the global CSI, and distributing them to their respective ETs for fully cooperative power transmission. Note that different from CoMP in wireless communication systems, where the user messages need to be shared among the cooperating base stations (BSs), the information exchanged among the  ETs mainly constitute their respective CSI. Thus, CoMP WPT systems in general have more relaxed requirement on the backhaul links than their communication counterparts. Note that CoMP WPT provides the performance upper bound for practical WPT systems with limited or no cooperation of ETs.

\subsubsection{Locally Coordinated WPT} For large WPT systems, it would be quite  challenging, if not impossible, for all ETs to fully cooperate. In this case, a more viable approach is to employ locally coordinated WPT, where each ER is locally served by $J'<J$ cooperating ETs. There are in general two approaches for locally coordinated WPT. With the {\it ET-oriented} approach, the $J$ ETs are partitioned into $G$ clusters with the $g$th cluster consisting of $J_g$ ETs, $g=1,\cdots, G$, and $\sum_{g=1}^G J_g=J$. As such, all the $J_g$ ETs within the same cluster $g$ will serve a subset of the ERs jointly. In contrast, with the more flexible {\it ER-oriented} approach, each of the $K$ ER is flexibly associated to a subset (in general different) of ETs based on certain criterion, such as the distance with the ET, as illustrated in Fig.~\ref{F:MultiUserWPT}. As such, two ERs with ER-oriented approach may have partially overlapped serving ETs,  in contrast to either identical or non-overlapping serving ETs in the ET-oriented approach. It is interesting to note that the channel reciprocity-based WPT techniques, such as the reverse-link training and retrodirective scheme, have the intrinsic capability to enable ER-oriented locally coordinated WPT. Specifically, with the reverse link pilot isotropically transmitted  from each ER, those nearby ETs would receive high pilot power and thus are more likely to be associated for cooperative power transmission to the ER.


\subsubsection{Single-ET WPT} For low-complexity WPT system, each ER is only served  by one single ET (e.g., the ET that has the best channel with it). This can be viewed as an extreme case of the locally coordinated WPT architecture with $J_g=1$, $\forall g$ and $G=J$, and hence essentially requires no coordination among the ETs. 

\subsection{Power Region Characterization}\label{sec:PowerRegion}
In this subsection, we derive the performance upper bound of multi-user WPT system by characterizing the power region of the CoMP WPT scheme for single-band systems, i.e., $N=1$ in \eqref{eq:xm}. Denote by $\mathbf x_j(t) \in \mathbb{C}^{M_t\times 1}$ the baseband equivalent energy-bearing signal sent by ET $j$, $j=1,\cdots, J$, and $\mathbf S_j\triangleq \mathbb{E}[\mathbf x_j(t)\mathbf x_j^H(t)]$ the corresponding transmit covariance matrix. We thus have $\tr(\mathbf S_j)\leq P_{\rf, j}^t$, $\forall j$, with $P_{\rf, j}^t$ denoting the transmit power limit at ET $j$. By assuming narrow-band channels, the equivalent baseband signal received at the ERs (with the noise ignored) can be expressed as
\begin{align}
\mathbf y_k(t) = \sum_{j=1}^J \mathbf H_{kj}\mathbf x_j(t), \ k=1,\cdots, K,\label{eq:yk}
\end{align}
where $\mathbf H_{kj} \in  \mathbb{C}^{M_r \times M_t}$ denotes the MIMO channel from ET $j$ to ER $k$, $k=1,\cdots, K$, $j=1,\cdots, J$.

%

For fully coordinated ETs, $\{\mathbf x_j(t)\}_{j=1}^J$ can be jointly designed to achieve the optimal performance, and hence they are correlated with each other in general. Let $\mathbf x(t)\triangleq \left[\mathbf x_1^T(t),\cdots \mathbf x_J^T(t)\right]^T \in \mathbb{C}^{JM_t\times 1}$ be the concatenated vector denoting the signal transmitted by all the $J$ ETs, and $\mathbf S\triangleq \mathbb{E}[\mathbf x(t) \mathbf x^H(t)]$ be the covariance matrix of $\mathbf x(t)$. The per-ET power constraint $\tr(\mathbf S_j)\leq P_{\rf,j}^t$ can then be equivalently expressed as
\begin{align}
\tr(\mathbf E_j \mathbf S)\leq P_{\rf,j}^t, \ j=1,\cdots, J,
\end{align}
where $\mathbf E_j\in \mathbb{C}^{JM_t\times JM_t}$ is a diagonal matrix with its diagonal elements given by
\begin{align}
[\mathbf E_j]_{mm}=\begin{cases}
1, \ &  (j-1)M_t+1 \leq m\leq jM_t\\
0, \ & \text{otherwise}.
\end{cases}
\end{align}
The received signal \eqref{eq:yk} can be equivalently expressed as $\mathbf y_k(t)=\mathbf H_k \mathbf x(t)$, $\forall k$, where
$\mathbf H_k\triangleq \left[\mathbf H_{k1},\cdots,\mathbf H_{kJ} \right]\in \mathbb{C}^{M_r\times JM_t}$ denotes the concatenated channel matrix associated with ER $k$. As a result, the received RF power $Q_k$ can be written as
\begin{align}
Q_k=\mathbb{E}[\|\mathbf y_k(t)\|^2]=\tr(\mathbf H_k^H \mathbf H_k \mathbf S), \ k=1,\cdots, K.
\end{align}

Different from the single-user WPT system, the design for multi-user WPT systems in general involves trade-offs in maximizing the transferred power to different users. In this case, the ETs can be optimally designed to maximize the {\it power region}, denoted by $\mathcal C$, which is defined as the set of all achievable power-tuples $(Q_1,\cdots, Q_K)$. Mathematically, we define
\begin{align}
\mathcal C = \bigcup_{\begin{substack}{\mathbf S\succeq \mathbf 0\\ \tr(\mathbf E_j \mathbf S)\leq P_{\rf,j}^t, \forall j}\end{substack}} \left \{(Q_1,\cdots, Q_K): Q_k\leq \tr(\mathbf H_k^H \mathbf H_k \mathbf S), \forall k \right\}.
\end{align}

Of particular interest is the Pareto boundary of the power region $\mathcal C$, which is defined as the power-tuples at which it is impossible to increase the received power of one ER without reducing that of the others. Similar to the capacity region in multi-user communication systems, the power region Pareto boundary  for multi-user WPT systems can be characterized via the {\it weighted-sum-power maximization} (WSPMax) approach or the {\it power-profile} approach, as explained in the following.

 With the WSPMax method, for each given weight vector $\boldsymbol \mu=[\mu_1,\cdots, \mu_K]^T$ for the $K$ ERs, with $\mu_k\geq 0$ and $\sum_{k=1}^K \mu_k=1$, the corresponding point on the Pareto boundary of the power region is determined by solving the following WSPMax problem,
 \begin{equation}\label{eq:ParetoCharacterizationWSPMax}
\begin{aligned}
\underset{\mathbf S}{\max} \ & \sum_{k=1}^K \mu_k \tr(\mathbf H_k^H \mathbf H_k \mathbf S) \\
\text{s.t. } \ &  \tr(\mathbf E_j \mathbf S)\leq P_{\rf,j}^t, \ \forall j=1,\cdots, J,\\
& \mathbf S\succeq \mathbf 0.
\end{aligned}
\end{equation}
Problem \eqref{eq:ParetoCharacterizationWSPMax} is a semidefinite programming (SDP), which is convex and can be efficiently solved by the standard convex optimization techniques or existing software toolbox such as CVX \cite{227}. Moreover, it is not difficult to show that the objective of problem \eqref{eq:ParetoCharacterizationWSPMax} is equivalent to that of the single-user WPT problem with an equivalent MIMO channel $\bar{\mathbf H}\in \mathbb{C}^{M_r \times JM_t}$ from the $J$ ETs to an auxiliary user, with $\bar{\mathbf H}^H \bar{\mathbf H} = \sum_{k=1}^K \mu_k \mathbf H_k^H \mathbf H_k$. In particular, for the special single-ET case, i.e., $J=1$, problem \eqref{eq:ParetoCharacterizationWSPMax} reduces to \eqref{P:SingleUser} with $N=1$, where the optimal solution is given by the dominating eigenbeam transmission over the effective channel $\bar{\mathbf H}$. Note that for power region constituting hyper-plane Pareto boundaries, the WSPMax approach only obtains those vertex points on the Pareto boundary, where time sharing is in general needed to attain the inner points on the boundary.

 On the other hand, with the {\it power-profile} method \cite{423}, the Pareto boundary of $\mathcal C$ can be characterized by solving the following optimization problem with any given power profile vector $\boldsymbol \alpha=(\alpha_1, \cdots \alpha_K)$ for the $K$ ERs,
\begin{equation}\label{eq:ParetoCharacterization}
\begin{aligned}
\underset{\mathbf S, Q}{\max} \ & Q \\
\text{s.t. } \ &  \tr(\mathbf H_k^H \mathbf H_k \mathbf S)\geq \alpha_k Q, \ \forall k=1,\cdots,K,\\
& \tr(\mathbf E_j \mathbf S)\leq P_{\rf,j}^t, \ \forall j=1,\cdots, J,\\
& \mathbf S\succeq \mathbf 0.
\end{aligned}
\end{equation}
where $\alpha_k\geq 0,  \forall k$ and $\sum_{k=1}^K \alpha_k=1$. 
Similar to \eqref{eq:ParetoCharacterizationWSPMax}, problem \eqref{eq:ParetoCharacterization} is also an SDP, which is convex and hence can be efficiently solved by, e.g., CVX \cite{227}. 

Denote by $\mathbf S^\star$ the optimal solution to problem \eqref{eq:ParetoCharacterization}. As the number of ERs $K$ becomes large, we have $d^\star\triangleq \mathrm{rank} (\mathbf S^\star)>1$ in general \cite{822}, i.e., more than one energy beams are needed for balancing the received energy among different ERs. In \cite{822}, an alternative design based on single-beam energy beamforming with time sharing transmission is proposed for the setup with a single ET, i.e., $J=1$, which is able to achieve the same optimal WPT performance as the multi-beam transmission. Specifically, for $J=1$, we must have $\tr(\mathbf S^\star)=P_{\rf,1}^t$, i.e., full power should be used at the optimal solution to problem \eqref{eq:ParetoCharacterization}. Let the eigenvalue decomposition of the optimal covariance matrix $\mathbf S^\star$ in the multi-beam transmission be expressed as $\mathbf S^\star = \sum_{i=1}^{d^\star} \lambda_i^\star \mathbf w_i^\star \mathbf w_i^{\star H}$, with $\lambda_i^\star>0$ and $\mathbf w_i^\star$ being the $i$th eigenvalue and the corresponding eigenvector, respectively. We then have $\sum_{i=1}^{d^\star} \lambda_i^\star =P_{\rf,1}^t$. With the proposed single-beam and time-sharing strategy in \cite{822}, each WPT transmission block is partitioned into $d^\star$ intervals, with the $i$th interval taking a fractional duration of $0< \lambda_i^\star/P_{\rf,1}^t< 1$. 
At the $i$th interval, the ET applies the single-beam energy transmission with beamforming vector $\mathbf w_i^\star$ with full power $P_{\rf,1}^t$. As a result, the average received power for ER $k$ during each block can be obtained as
\begin{align}
\sum_{i=1}^{d^\star} \frac{\lambda_i^\star}{P_{\rf,1}^t} \tr\left(\mathbf H_k^H \mathbf H_k (P_{\rf,1}^t \mathbf w_i^\star \mathbf w_i^{\star H})\right)=\tr(\mathbf H_k^H \mathbf H_k \mathbf S^\star), \ \forall k.
\end{align}
In other words, the newly designed  single-beam transmission with time sharing achieves the same energy performance for all ERs as the optimal multi-beam transmission with  $\mathbf S^\star$, but requires only single-beam transmission at each interval, thus simplifying the power signal design at the ET. 



\subsection{Numerical Results}
\begin{figure}
\centering
\includegraphics[scale=0.55]{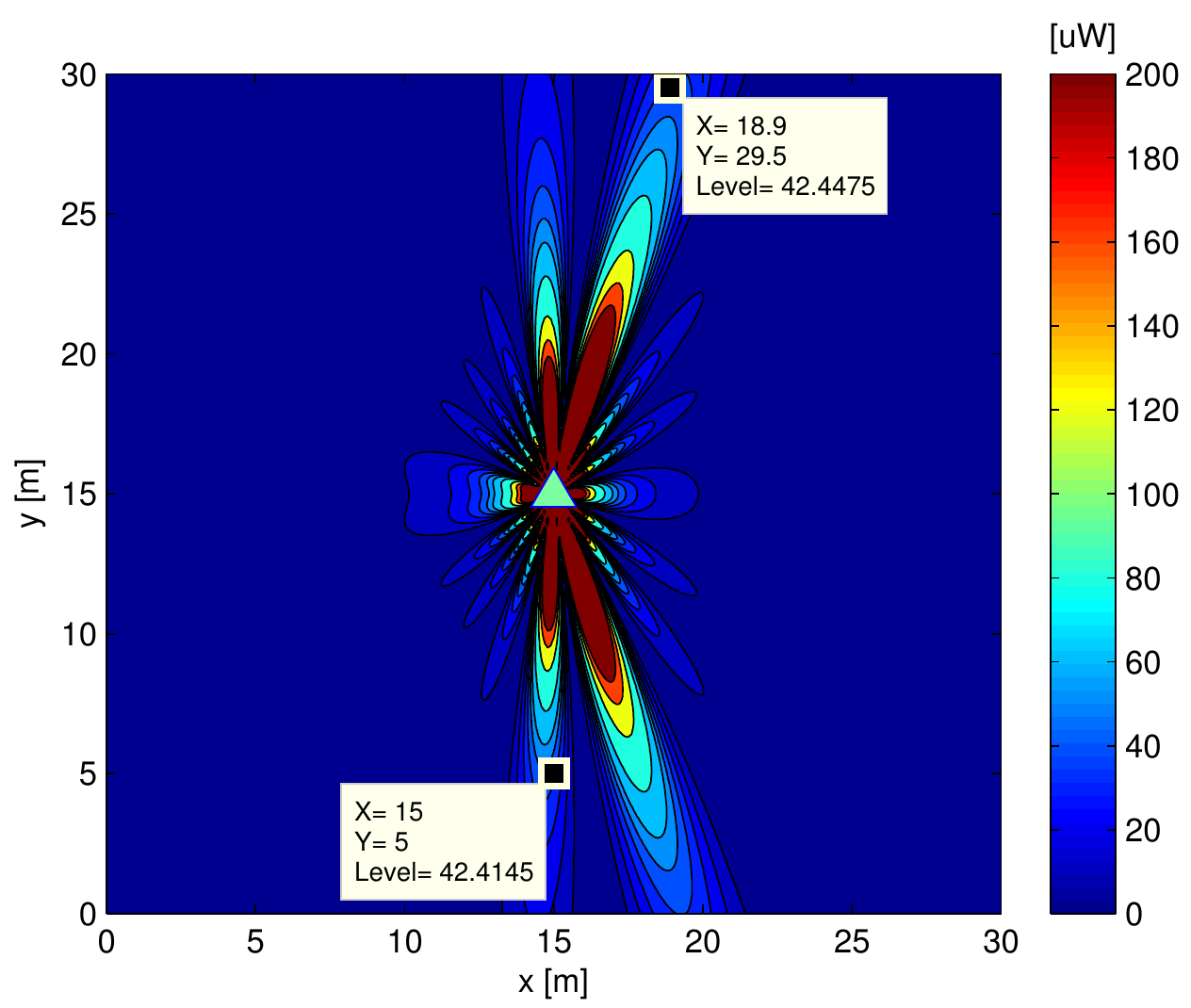}\\
(a)\\
\includegraphics[scale=0.55]{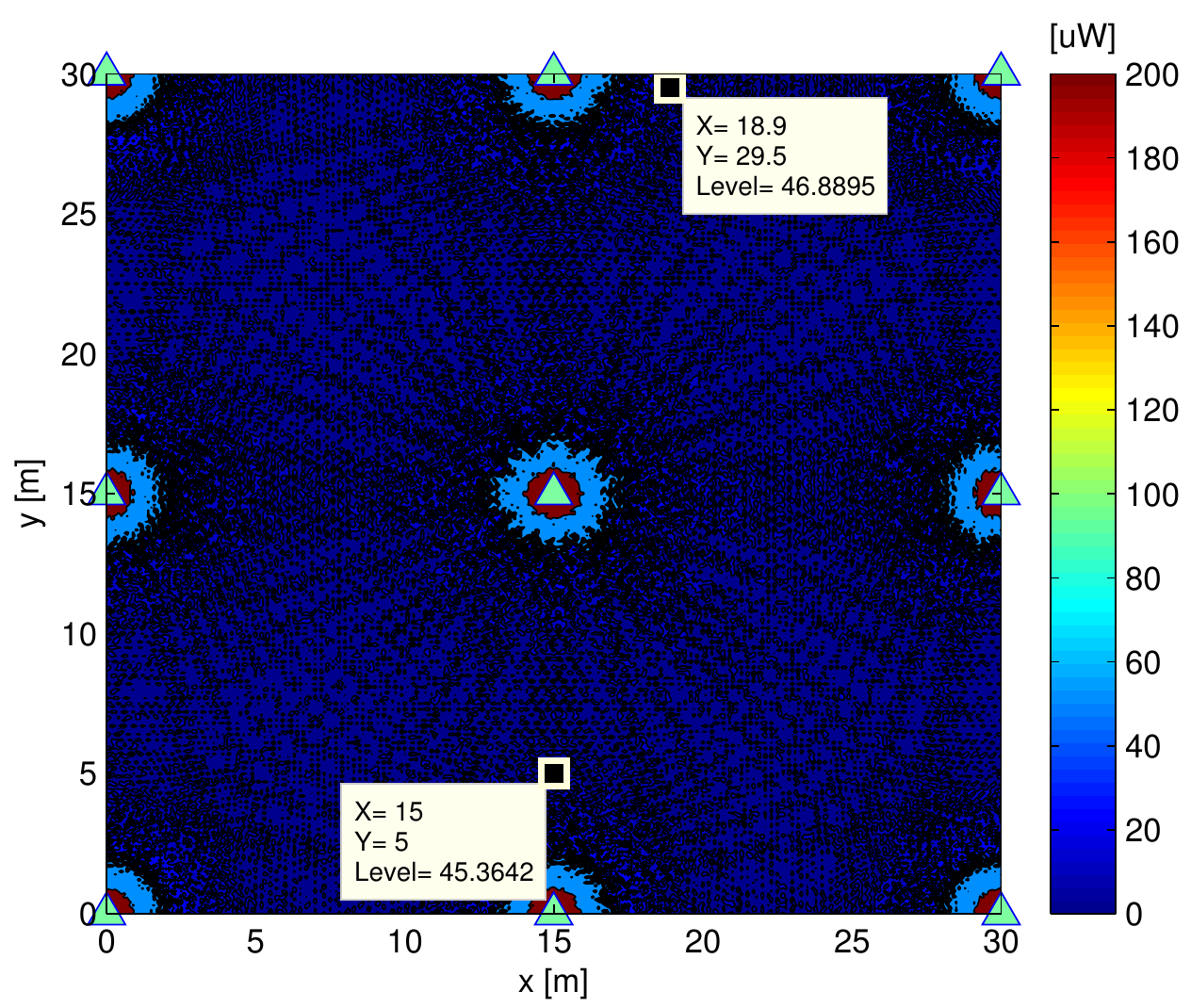}\\
(b)
\caption{Spatial power distribution with max-min WPT for: (a) {\it co-located antenna system} with a single ET equipped with an ULA of $M_t=9$ antennas; (b) {\it distributed antenna system} with $J=9$ single-antenna ETs that are equally spaced. Triangle and square represent ET and ER, respectively.}\label{F:PowerDisComparison}
\end{figure}

 For illustration, we consider a WPT system that serves a square area of size $30$m $\times$ $30$m, as shown in Fig.~\ref{F:PowerDisComparison}. We compare the {\it co-located} and {\it distributed} antenna systems \cite{858}. In the  co-located antenna system, a single ET with an $M_t$-element uniform linear array (ULA) is deployed at the center of the serving area with coordinate (15m, 15m), as shown in Fig.~\ref{F:PowerDisComparison}(a). We assume that $M_t=9$ and the ULA is oriented along the x-axis. In contrast, for the distributed antenna system, we assume that $J=9$ single-antenna ETs are equally spaced in the region, as shown in Fig.~\ref{F:PowerDisComparison}(b). We consider two single-antenna ERs that are located at (15m, 5m) and (18.88m, 29.49m), respectively, which correspond to a distance of 10m and 15m from the ET in the co-located antenna system. We assume that the channels between the ETs and ERs are dominated by LoS links and the carrier frequency is $915$MHz. The total transmit power of both systems is 2W or 33dBm, which needs to be equally shared by the $9$ ETs in the distributed antenna system. Moreover, for the distributed antenna system, we assume that the CoMP-based WPT strategy is applied.

\begin{figure}
\centering
\includegraphics[scale=0.5]{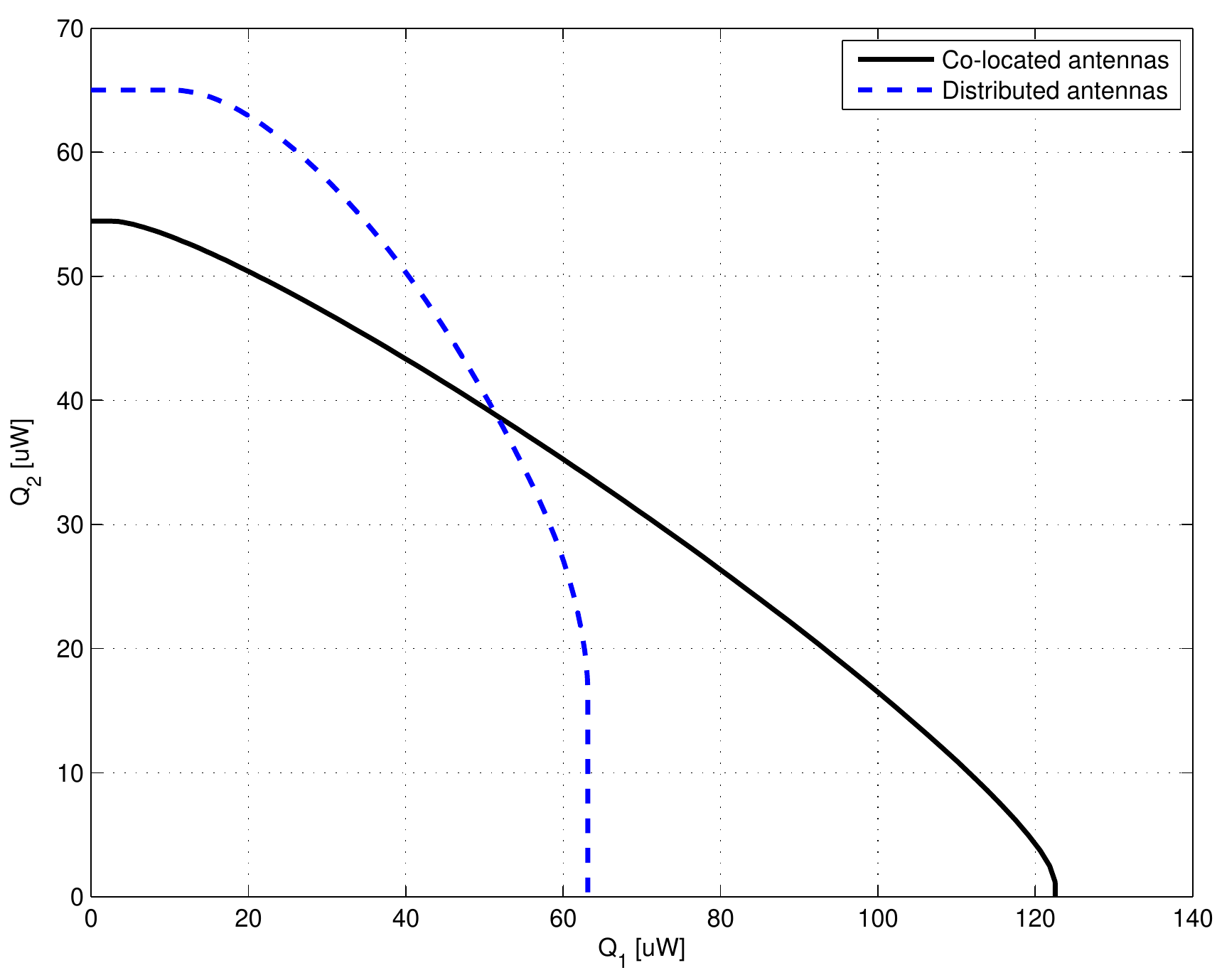}
\caption{Power region for two-user WPT system with co-located versus distributed antenna systems.}\label{F:PowerRegion}
\end{figure}

Fig.~\ref{F:PowerDisComparison} shows the spatial power distribution of the two WPT systems when the transmission is optimized for maximizing the minimum (max-min) received power by the two ERs, i.e., by solving problem \eqref{eq:ParetoCharacterization} with $\alpha_1=\alpha_2=1/2$. It is observed from Fig.~\ref{F:PowerDisComparison}(a) that for the co-located antenna system, the power is mainly beamed towards the directions (the actual direction and its symmetrical one over the x-axis) where the two ERs are located. In contrast, with the distributed antenna system as shown in Fig.~\ref{F:PowerDisComparison}(b), no evident energy focusing direction is observed and the power is more evenly distributed in space compared to the co-located system. It is also observed that the distributed system achieves a slightly higher max-min power than the co-located system (45.4$\mu$W versus 42.4$\mu$W), thanks to the reduced distance between ER$_2$ and its nearest ET in the distributed case. Fig.~\ref{F:PowerRegion} compares the complete power regions of the two WPT systems. It is observed that  the distributed antenna system achieves higher  maximum power for ER$_2$, but at the cost of reduced maximum power for ER$_1$. In other words, by placing antennas at different locations, the distributed system may potentially mitigate the near-far problem in the co-located system, and hence is expected to achieve more fair performance between the ERs.


\subsection{Extension and Future Work}
The power region characterization for multi-user multi-antenna WPT in frequency-selective channels deserves further studies. In particular, due to the unique channel frequency responses, different ERs may prefer the wireless energy to be transmitted over different frequencies. This thus provides another degree of freedom, in addition to the spatial beamforming, to achieve different trade-offs on the Pareto boundary of the power region. Moreover, besides power region that characterizes the long-term average power trade-off for ERs,  another useful performance metric for multi-user WPT systems is the {\it energy outage} region, which specifies the outage probability trade-off among the ERs with their given short-term energy targets, and thus is a more appropriate design criterion for delay-sensitive charging applications in fading channels. The characterization of the energy outage region deserves more in-depth studies. Moreover, the channel acquisition for multi-user MIMO WPT systems  in both frequency-flat and frequency-selective channels is a promising direction for further research.  Note that for revere-link based training in multi-user WPT systems, the optimal  training design needs to tackle the so-called ``doubly near-far'' problem \cite{528}, where a far ER from the ET suffers from higher propagation loss than a near ER for both reverse-link channel training and forward-link energy transmission. Furthermore, for WPT networks to be scalable, the transmit optimization and channel learning need to be implemented in a distributed manner \cite{777}, \cite{841}, with limited or no signaling overhead among different nodes. For large-scale WPT networks, stochastic geometry is a useful tool for performance analysis and optimization to draw useful insights \cite{859}, \cite{842}. How to optimally deploy the ETs to minimize their number (cost) to cover a group of distributed wireless nodes to satisfy their energy and communication demands is also an interesting problem for investigation \cite{860}.

\section{Waveform Design with Non-Linear Energy Harvesting Model}\label{Section_NL}
The major challenge for far-field WPT is to find ways to increase the DC power level at the output of the rectenna without increasing the transmit power, and for devices located tens to hundreds of meters away from the transmitter. To that end, the energy beamformer was shown to increase the RF-to-RF transmission efficiency $e_2$. At the receiver side, the vast majority of the technical efforts in the literature to increase the RF-to-DC conversion efficiency $e_3$ have been devoted to the design of efficient rectennas, a.o. \cite{502}. It therefore appears that an efficient design of WPT system would consist in an energy beamformer designed so as to maximize $e_2$ and an efficient rectenna that maximizes $e_3$. However, this may not be as efficient as expected and could lead to suboptimal designs. Recall indeed that the main assumption on the energy beamformer design that maximizes $e_2$ is that $e_3$ is fixed and therefore independent of the input signal power and shape to the rectenna. This is actually true only for very small input power, as it will appear clearer in this section. Indeed, the RF-to-DC conversion efficiency $e_3$ of the rectenna is in general not only a function of the rectenna design but also of its input waveform. This calls for an entire link optimization where the transmit waveform (including energy beamformer) is optimized to maximize $e_2 \times e_3$ for a given rectenna design, and not only $e_2$ \cite{Clerckx:2015}, \cite{Clerckx:2016b}. This would lead to a radically different system design than the one obtained in Sections~\ref{sec:Pt2Pt} and \ref{sec:WPTNetwork} and is discussed in detail in this section.

\subsection{Effect of Non-Linearity on RF-to-DC Conversion Efficiency}\label{NL_e3}

In order to get some insight into the effect of the rectifier non-linearity on $z_{DC}$ given in \eqref{polynomialSeries_DC_4}, we consider in the sequel two toy examples, the first one over a frequency-flat channel and the second one over a frequency-selective channel.
\par Let us first consider a SISO (single-input single-output) WPT system with $N$ sinewaves equally spaced, i.e.\ $f_n=f_1+(n-1) \Delta_f$, with given $\Delta_f> 0$ and assume a frequency-flat channel such that the channel frequency response $h_n=1$ $\forall n$. We also assume that the weights $s_n$ are deterministic, real and subject to the transmit power constraint $\sum_n s_n^2=P_{\textnormal{rf}}^t$. With such normalization, $P_{\textnormal{rf}}^r=P_{\textnormal{rf}}^t$. From (3), the received signal can be written as $y(t)=\sqrt{2}\Re\big\{\sum_{n=1}^{N}s_n e^{\jmath 2 \pi f_n t}\big\}=\sqrt{2}\sum_{n=1}^{N}s_n \cos\left(2 \pi f_n t\right)$, i.e., as the sum of $N$ in-phase sinewaves, each with a magnitude $s_n$. Plugging $y(t)$ into \eqref{polynomialSeries_DC_4}, we obtain
\begin{equation}\label{i_out_FF}
z_{DC}= k_2 R_{ant} P_{\textnormal{rf}}^r+\frac{3 k_4}{2} R_{ant}^2 F
\end{equation}
where
\begin{equation}\label{F}
F=\sum_{\mycom{n_0,n_1,n_2,n_3}{n_0+n_1=n_2+n_3}} s_{n_0}s_{n_1}s_{n_2}s_{n_3}.
\end{equation}
We note from \eqref{i_out_FF} that the second-order term $k_2 R_{ant} P_{\textnormal{rf}}^r$ is independent of the number of sinewaves $N$ and the power allocation strategy in such a frequency-flat channel. This is inline with the discussion in Section \ref{sec:EnergyBF} on energy beamforming. On the other hand, the fourth-order term is responsible for the non-linear behavior of the diode since it is a function of terms expressed as the product of contributions from different frequencies. Contrary to the second-order term, the fourth-order term is heavily influenced by $N$ and the choice of the power allocation strategy. Though not optimal, let us consider a uniform power allocation across all frequencies, i.e.\ $s_n=\sqrt{P_{\textnormal{rf}}^t}/\sqrt{N}$. Since there are $N\left(2N^2+1\right)/3$ terms in the sum of \eqref{F}, we get with a uniform power allocation that
\begin{align}
z_{DC}&= k_2 R_{ant} P_{\textnormal{rf}}^r+k_4 R_{ant}^2 \frac{2N^2+1}{2N}{P_{\textnormal{rf}}^r}^2,\nonumber\\
&\stackrel{N\nearrow}{\approx}k_0+k_2 R_{ant} P_{\textnormal{rf}}^r+k_4 R_{ant}^2 N {P_{\textnormal{rf}}^r}^2.\label{UP_flat_N}
\end{align}
Remarkably, \eqref{UP_flat_N} highlights that $z_{DC}$, and therefore $i_{out}$, linearly increase with $N$ in frequency-flat channels and such an increase originates from the non-linearity of the rectifier as it only appears in the fourth-order term. Hence, while there is no benefit in allocating power over multiple sinewaves with the linear model, and simply transmitting over a single sinewave would be sufficient, the non-linear model clearly highlights a completely different strategy where power should be transmitted over multiple sinewaves. Interestingly, this stragegy is in agreement with various RF experiments \cite{749}, \cite{Trotter:2009,Boaventura:2011,Collado:2014} where the benefits of allocating power over multiple sinewaves have been demonstrated experimentally.
More generally, it is shown in \cite{Clerckx:2016b} that the linear increase of $z_{DC}$ with $N$ holds both in frequency-flat and frequency-selective channels. However, while it is achievable without CSIT in frequency-flat channels, CSIT is required to achieve such a scaling law in frequency-selective channels. It is also shown in \cite{Clerckx:2016b} that in the presence of multisine and multiple transmit antennas, the fourth-order term scales as $N M_t^2$, suggesting that any increase of $z_{DC}$ by a factor 2 requires either increasing the number of sinewaves ($N$) by a factor 2 for a fixed number of transmit antennas ($M_t$) or increasing the number of transmit antennas by a factor $\sqrt{2}$ for a fixed number of sinewaves.

\par
\begin{figure}
\centerline{\includegraphics[width=\columnwidth]{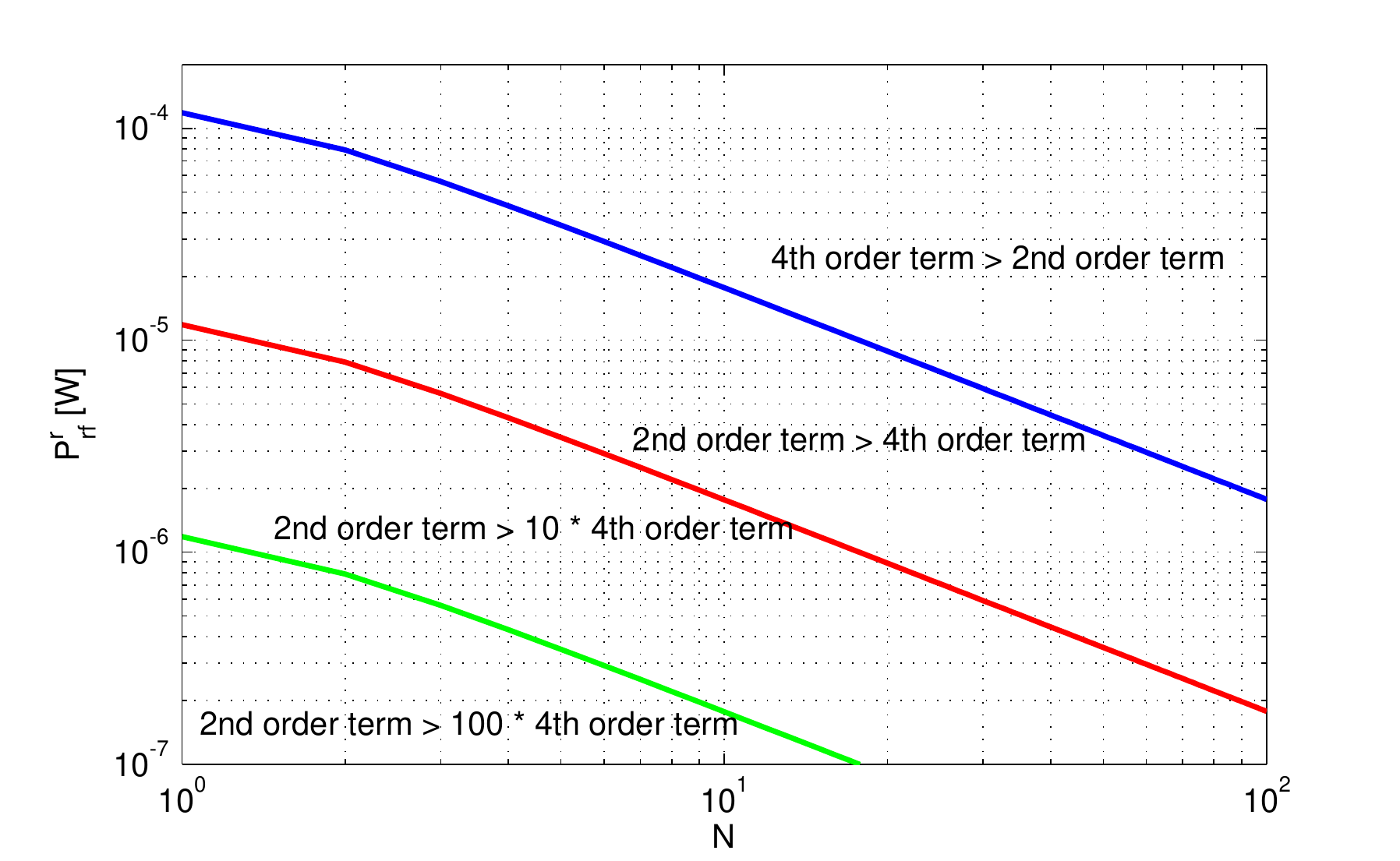}}
  \caption{Linear vs non-linear regime.}
  \label{2nd_vs_4th}
\end{figure}

 Scaling law \eqref{UP_flat_N} enables to characterize the strength of the fourth-order term versus the second-order term. Specifically, the second-order term is $G$ times larger than the fourth-order term if
\begin{equation}
P_{\textnormal{rf}}^r\leq \frac{k_2}{k_4}\frac{1}{R_{ant}}\frac{1}{N}\frac{1}{G}.
\end{equation}
Assuming $i_s=5 \mu A$, a diode ideality factor $n_f=1.05$ and $v_t=25.86 mV$, typical values are $k_2=0.0034$, $k_4=0.3829$ and $R_{ant}=50\Omega$, which lead to $\frac{k_2}{k_4}\frac{1}{R_{ant}} = 1.776\times10^{-4}$. This is further illustrated in Fig.~\ref{2nd_vs_4th}. We note that for an average input power of $10\mu W$ ($-20$ dBm), the nonlinearity is not negligible compared to the second-order term for most $N$. For an average input power of $1 \mu W$ ($-30$ dBm), the nonlinearity is negligible for $N$ smaller than roughly 20. Note however that $-30$ dBm is actually very small for  state-of-the-art rectifiers.

\par The physical intuition behind the linear increase of the DC current with $N$ is as follows. For $N$ in-phase sinewaves over a frequency-flat channel, as $N$ increases, the time domain waveform appears as a sequence of pulses with a periodicity equal to $1/\Delta_f$. The transmitter therefore concentrates the transmit power into a series of high energy pulses, each of which triggers the diode that then conducts and helps charging the output capacitor. Once a pulse has passed, the diode stops conducting and the capacitor is discharging. The larger $N$, the larger is the magnitude of the pulses and therefore the larger the output voltage at the time of discharge. This intuitively explains why a multisine waveform with high PAPR (Peak-to-Average Power Ratio) helps increasing the output DC power. Interestingly, experimental results in \cite{Collado:2014} have shown that waveforms with high PAPR, such as OFDM (orthogonal frequency division multiplexing), white noise, and chaotic signals, increase the RF-to-DC conversion efficiency. Nevertheless, we have to keep in mind that this observation holds for frequency-flat channels. In frequency-selective channels, the correlation between transmit PAPR and DC current/power decreases as the selectivity increases \cite{Clerckx:2016b}.

\par It is very important to recall that the linear increase with $N$ is based on multisine signals with deterministic weights. This is a key assumption. If the weights are pseudo-random due to e.g.\ modulation as in OFDM, the DC current at the output of the rectifier fluctuates due to the randomness of the information symbols carried by the modulated waveform \cite{Clerckx:2016c}. Assuming for instance that weights $s_n$ are i.i.d. CSCG distributed (following the capacity achieving Gaussian input distribution of an AWGN (additive white Gaussian noise) communication channel with average power constraint), the average DC current with a modulated waveform is modeled in \cite{Clerckx:2016c} by averaging out \eqref{polynomialSeries_DC} over the distribution of the input symbols. This has an important consequence that the linear increase with $N$ in the 4th-order term of \eqref{UP_flat_N} disappears with modulated waveform. On the other hand, the 2nd-order term is not affected even when the waveform is modulated. Hence, from a linear model perspective, modulated waveform (as OFDM) and deterministic multisine waveforms are equally suitable. On the other hand, the nonlinear model highlights that there is a clear benefit of using a deterministic multisine over a modulated (OFDM) waveform in WPT, with the scaling law of multisine significantly outperforming that of OFDM. This shows that due to the non-linearity of the rectifier and any pseudo-randomness (due to modulation), a modulated waveform is less efficient than a deterministic multisine waveform for WPT. Further discussion and comparisons between deterministic multisine and OFDM waveforms can be found in \cite{Clerckx:2016c}.

\par Let us now look at a SISO WPT but over a frequency-selective channel. We assume for simplicity $N=2$ and assume again real frequency domain channel $h_n$ and real (and deterministic) coefficient $s_n$. The received signal at the input of the rectenna now writes as $y(t)=\sqrt{2}\Re\big\{\sum_{n=1}^{2}s_n h_n e^{\jmath 2 \pi f_n t}\big\}=\sqrt{2}\sum_{n=1}^{2}s_n h_n \cos\left(2 \pi f_n t\right)$. Plugging $y(t)$ into \eqref{polynomialSeries_DC_4}, we obtain
\begin{multline}
z_{DC}=\tilde{k}_2 \left(s_1^2 h_1^2+s_2^2 h_2^2\right) \\
+\tilde{k}_4  \left[\left(s_1^2 h_1^2+s_2^2 h_2^2\right)^2+ 2 s_1^2s_2^2 h_1^2 h_2^2\right]\label{diode_model_2_N2_1}
\end{multline}
where $\tilde{k}_2=k_2 R_{ant}$ and $\tilde{k}_4=3 k_4 R_{ant}^2/2$. We note that $z_{DC}$ is a function of the term $s_1^2 h_1^2+s_2^2 h_2^2$, whose maximization subject to the sum power constraint $s_1^2+s_2^2\leq P_{\textnormal{rf}}^t$ would lead to a single-sinewave strategy, i.e.\ allocating all the power to sinewave 1 if $h_1^2>h_2^2$ and to sinewave 2 otherwise, as discussed in Section~\ref{sec:EnergyBF}. However the presence of the term $2 s_1^2s_2^2 h_1^2 h_2^2$ suggests that such a single-sinewave strategy is in general sub-optimal for the maximization of $z_{DC}$. This can be shown by writing the Lagrangian and finding all the stationary points. We can find three valid stationary points $(s_1^2,s_2^2)$ (such that $0 \leq s_1^2\leq P_{\textnormal{rf}}^t$ and $0 \leq s_2^2\leq P_{\textnormal{rf}}^t$) given by
$(P_{\textnormal{rf}}^t,0)$, $(0,P_{\textnormal{rf}}^t)$ and $(s_1^{\star 2},s_2^{\star 2})$ with $s_1^{\star 2}>0$, $s_2^{\star 2}>0$. For given $h_1$, $h_2$, the global optimum strategy is given by one of those three stationary points. The first two points correspond to a single-sinewave strategy, i.e.\ allocating the full transmit power to sinewave 1 or 2, respectively. This strategy is optimal if $h_1^2$ is sufficiently larger than $h_2^2$ or inversely. However, when the channel is getting more frequency-flat, i.e.\ $h_1^2 \approx h_2^2$, the optimal strategy would allocate power to the two sinewaves and the single-sinewave strategy is suboptimal. A more detailed illustration of $z_{DC}$ as a function of the channel states and further derivation of $s_1^{\star}$ and $s_2^{\star}$ can be found in \cite{Clerckx:2016b}.
\par This example highlights that due to the non-linear behaviour of the rectifier, it may be preferable depending on the channel states to allocate power over two sinewaves so as to maximize $z_{DC}$ even though the maximization of $e_2$ would favour a single-sinewave strategy. In other words, the single sinewave strategy would always maximize $e_2$ but could be inefficient from an $e_3$ maximization perspective, such that a better strategy would be to allocate power over two sinewaves so as to maximize the output DC current (or power) and hence maximize the entire link efficiency ($e_2 \times e_3$).

\par The results in this subsection, though based on very simple scenarios, highlight that depending on the CSI, the transmission waveform should be adapted if we aim at maximizing the output DC
power (and the entire link efficiency). Acquiring CSIT at the transmitter so as to design adaptive waveform is therefore essential for the design of efficient WPT. Moreover, they also show the benefits of allocating power over multiple sinewaves, which is in sharp contrast with the strategy originating from the linear model and the maximization of $e_2$ only. This multi-band frequency allocation is reminiscent of multi-band wireless communication where the maximization of the achievable rate commonly requires allocating power over multiple frequency bands or spatial eigenmodes (at least at sufficiently high signal-to-noise ratio (SNR)). Overall, the observation highlights the potential of optimizing multisine waveforms and the importance of modeling and ``exploiting'' the non-linearity of the rectifier.
\par In the sequel, we discuss the design and optimization of multisine waveform to maximize the output DC power.

\subsection{Waveform Design}\label{multisine_waveform_design}
Assuming the CSI (in the form of frequency response $h_{n}$) is known to the transmitter, we aim at finding $\mathbf{s}=\left[s_1,\ldots,s_N\right]$, the vector of complex weights $s_n$ over $N$ frequencies, which maximizes $z_{DC}$ in \eqref{z_DC_def} for any $n_o\geq 2$. Importantly, we assume deterministic weights $s_n$. Let us assume $M_t=1$ and $M_r=1$. The multisine waveform design problem can therefore be written as
\begin{equation}\label{P2}
\max_{\mathbf{s}} \hspace{0.2cm} z_{DC}(\mathbf{s}) \hspace{0.3cm} \textnormal{subject to} \hspace{0.3cm} \left\|\mathbf{s}\right\|_F^2\leq P_{\textnormal{rf}}^t
\end{equation}
where $z_{DC}$ can be analytically expressed after plugging the received signal $y(t)$ of \eqref{eq:yit} into \eqref{z_DC_def}. The expressions are omitted but can be found in \cite{Clerckx:2016b}.
\par From \cite{Clerckx:2016b}, the optimal phases are given by $\textnormal{phase}\left(s_n\right)=\textnormal{phase}\left(h_n\right)$, i.e.\ the transmitter matches the phases of the channel on each frequency such that the multisine signal arrives in phase at the rectenna. Making use of those optimum phases, the optimum amplitudes result from a non-convex posynomial maximization problem which can be recast as a Reverse Geometric Program (GP) and solved iteratively using a successive convex approximation approach. This involves approximating (conservatively) the non-convex problem by a convex problem using the Arithmetic Mean-Geometric Mean (AM-GM) inequality and refining at each iteration the tightness of the approximation. The algorithm ultimately converges to a point fulfilling the KKT (Karush-Kuhn-Tucker) conditions of the original problem.
\par
\begin{figure}
\centerline{\includegraphics[width=\columnwidth]{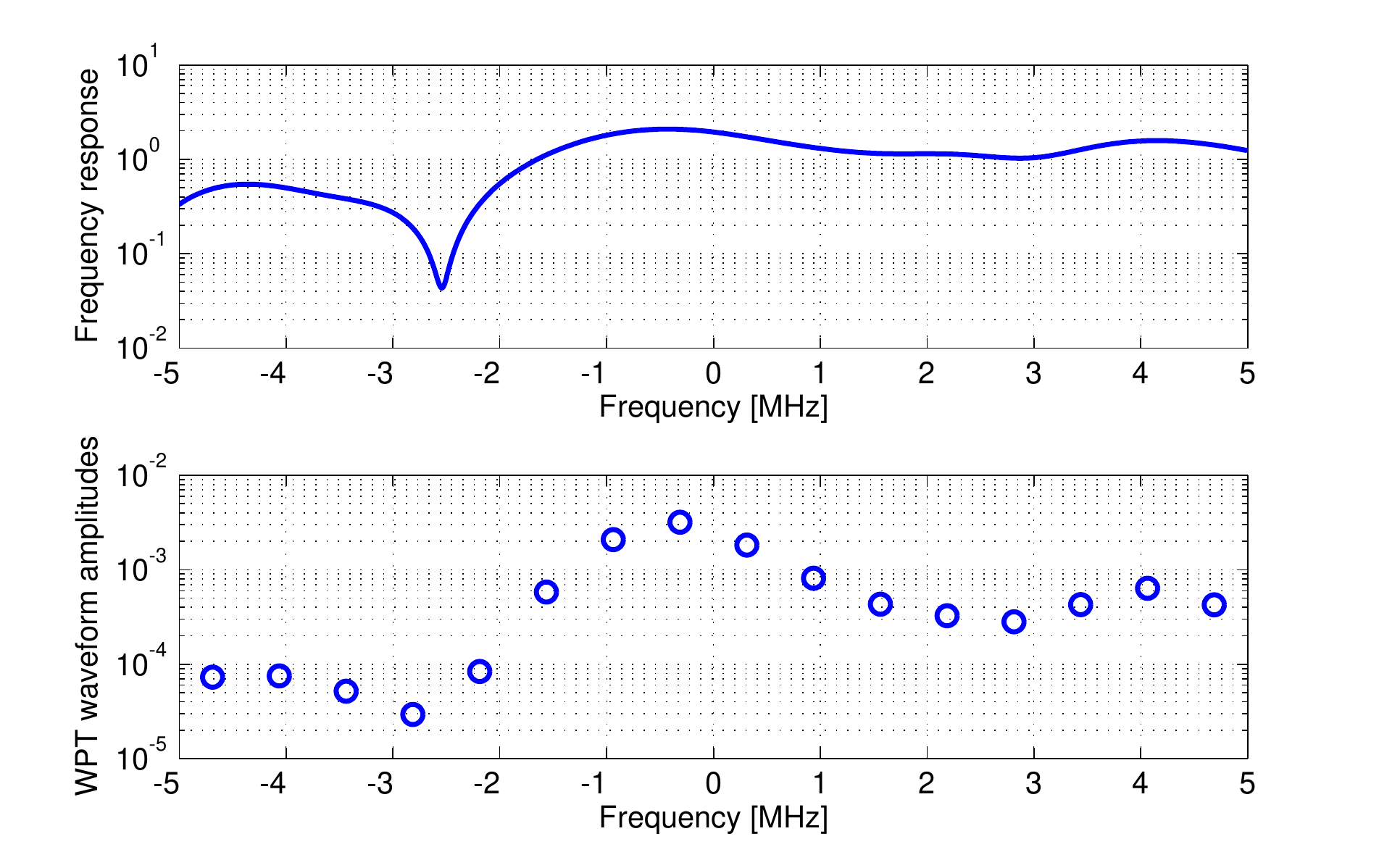}}
  \caption{Frequency response of the wireless channel and WPT waveform magnitudes ($N=16$) for 10 MHz bandwidth.}
  \label{Freq_response_channel1}
\end{figure}

 Fig.~\ref{Freq_response_channel1} provides some insights into the waveform optimization. We consider a frequency-selective channel whose frequency response is given in Fig.~\ref{Freq_response_channel1} (top), a transmit power of $-20$ dBm, $N=16$ sinewaves with a frequency gap fixed as $\Delta_f=B/N$ and $B=10$MHz. For such a channel realization assumed perfectly known (at each of those 16 frequencies) to the transmitter, the waveform has been optimized using the reverse GP algorithm assuming a 4th-order Taylor expansion ($n_o=4$). The magnitudes of the waveform on the 16 frequencies are displayed on Fig. \ref{Freq_response_channel1} (bottom). Interestingly, the optimized waveform has a tendency to allocate more power to frequencies exhibiting larger channel gains. Note that the 4th-order term is clearly not negligible in the objective function. Indeed if it was negligible, the entire transmit power would have been allocated to a single sinewave, namely the one among the 16 sinewaves corresponding to the strongest channel.

\par The waveform optimization can be extended to scenarios with multiple transmit antennas \cite{Clerckx:2016b}. A direct approach is to jointly optimize weights across space and frequency using the reverse GP approach. Interestingly, despite the presence of the non-linearity, such a joint optimization can be avoided by noting that the optimum $M_t$-dimensional weight vectors can be computed by first performing a MRT (maximal ratio transmission) beamformer on every frequency and then optimizing the power allocation across frequency based on the effective beamformed channel \cite{Clerckx:2016b}, \cite{Huang:2016b}. Hence an optimum $M_t$-dimensional weight vector on frequency $n$ can be written as
\begin{equation}\label{opt_weight}
\mathbf{s}_n=\sqrt{p_n} \frac{\mathbf{h}_n}{\left\|\mathbf{h}_n\right\|}.
\end{equation}
With \eqref{opt_weight}, the multi-antenna multisine WPT weight optimization is converted into an effective single-antenna multisine waveform optimization with the effective channel gain on frequency $n$ given by $\left\|\mathbf{h}_n\right\|$ and the amplitude of the $n$-th sinewave given by $\sqrt{p_n}$ (subject to $\sum_{n=1}^{N}p_n=P_{\textnormal{rf}}^t$). This enables to decouple space and frequency optimization by first designing the spatial (or energy) beamformer as a matched beamformer and then optimizing the frequency power allocation using the reverse GP approach. Interestingly, in the limit of a very large (infinite) number of transmit antennas, the design gets simpler. Indeed, the matched beamformer would induce channel hardening, therefore turning the frequency-selective channel into an effective flat-fading channel. Since a simple uniform power allocation $p_n=P_{\textnormal{rf}}^t/N$ works well in a SISO flat-fading channel, it can be leveraged in the large-scale multi-antenna regime to provide pretty convincing performance and a low waveform design complexity. Such a strategy would lead to a 4th-order term scaling as $N M_t^2$, as briefly described in Section \ref{NL_e3}.

\par The waveform optimization using reverse GP can also be extended to more complicated scenarios accounting for transmit PAPR constraints and for the presence of multiple rectennas \cite{Clerckx:2016b}. Since the transmit power amplifier (PA) efficiency decreases as the PAPR of the transmit waveform increases, it is important in some applications to identify how to optimize waveform subject to PAPR constraints. Problem \eqref{P2} can then be expanded by adding a per-antenna PAPR constraint ($m=1,\ldots,M_t$) as
\begin{align}\label{P2_PAPR}
\max_{\mathbf{s}} \hspace{0.3cm} &z_{DC}(\mathbf{s})\\
\textnormal{subject to} \hspace{0.3cm} &\left\|\mathbf{s}\right\|_F^2\leq P_{\textnormal{rf}}^t,\\
&PAPR_m\leq \eta, \forall m.\label{P2_2}
\end{align}
The problem can be expressed as a signomial (rather than a posynomial) maximization problem and can also be solved iteratively using successive convex approximation based on AM-GM. Recall that in frequency-flat channels the amount of collected energy is positively correlated with the PAPR of the transmitted power waveform. This creates a design challenge since high PAPR is detrimental from a PA perspective but beneficial from an energy collection perspective. Note however that in the presence of PAPR constraints, the decoupling properties between space and frequency does not hold anymore. Solving the spatial domain beamformer before optimizing the frequency power allocation would lead to a suboptimal solution.

\par The multi-rectenna scenario can refer to either a multi-user setup where each device is equipped with a rectenna or a point-to-point setup with a receiver equipped with multiple rectennas. The objective function can then be formulated as a weighted sum of DC current $z_{DC,u}$ at each rectenna ($u=1,\ldots,K$)
\begin{equation}\label{P2_MU}
\max_{\mathbf{s}} Z_{DC}(\mathbf{s})=\sum_{u=1}^K v_u z_{DC,u}(\mathbf{s}) \hspace{0.2cm}  \textnormal{s.t.} \hspace{0.2cm} \left\|\mathbf{s}\right\|_F^2\leq P_{\textnormal{rf}}^t.
\end{equation}
Here also, the optimization problem can be formulated as a signomial maximization problem. However, contrary to the single rectenna setup where the optimal phase could be first obtained and then the magnitudes optimized, in the multiple rectenna setup, the phase and magnitude are coupled. Formulating \eqref{P2_MU} as a signomial maximization problem requires an initial choice for the phase before the magnitudes can be optimized and there is no guarantee that this choice of phase is optimal.

\par The reversed GP approach to waveform optimization is powerful in that it can be applied to any order $n_o$ in the Taylor expansion but suffers from exponential complexity. This is problematic for large-scale waveform optimization in WPT system relying on a large number of transmit antennas, sinewaves and/or rectennas (i.e.,  Massive MIMO of WPT). This calls for a reformulation of the optimization problem by expressing the RF signal model in a compact form using a real-valued function of complex vector variables \cite{Huang:2016}, \cite{Huang:2016b}. The compact expression is essentially
a quartic function that in general still leads to NP-hard problems. To make the problem tractable, auxiliary variables are introduced and convex relaxations are used such that the quartic objective can be reduced to a non-convex quadratic constraint in an equivalent problem. Then, the non-convex constraint is linearized, and the equivalent problem is iteratively approximated. Following this, a variety of convex optimization techniques (e.g., successive convex approximation (SCA), rank reduction) can be used to solve the approximate problem. The waveform
optimization framework is derived for a single-user/rectenna WPT and is then generalized to multi-user/rectenna WPT systems. The objective function can be written as a weighted sum of DC current/voltage as in \eqref{P2_MU} but we also need to tackle the maximization of the minimum DC current/voltage among all rectennas in order to guarantee some fairness among users. Contrary to the single-rectenna scenario, the optimal design is obtained by a joint spatial domain beamformer and frequency power allocation in the multi-rectenna scenario.

\subsection{Performance Evaluations}\label{eval}

\begin{figure}
\centerline{\includegraphics[width=0.9\columnwidth]{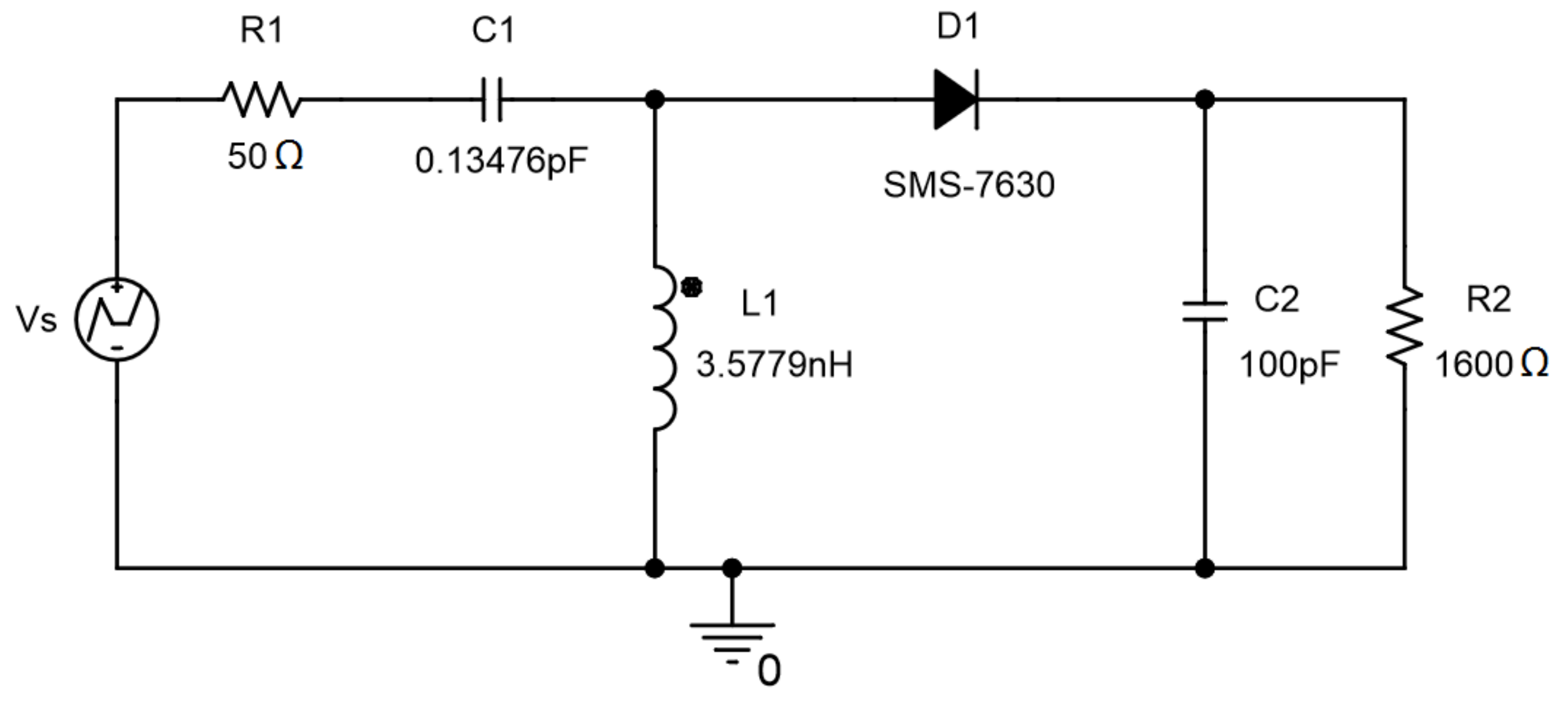}}
  \caption{Rectenna with a single diode and an L-matching network used for PSpice evaluations with $B=1$MHz.}
  \label{circuit}
\end{figure}

\par In this subsection, we evaluate the performance of the waveforms using the rectifier configuration of Fig. \ref{circuit}. The rectenna is optimized for a multisine input signal composed of 4 sinewaves centered around 5.18GHz with the bandwidth of 1MHz. The available RF power is $P_{in,av}=-20$dBm.  The input impedance of the rectifier $Z_{rect}$ is dominated by the diode impedance, which changes depending on the input power and the operating frequency. In order to avoid power losses due to impedance mismatch, the matching network design procedure is adapted for a multisine input signal of varying instantaneous power. The matching is done by iterative measurements of $Z_{rect}$ at the 4 sinewave frequencies using ADS Harmonic Balance and performing conjugate matching of average $\bar Z_{rect}$ to $R_{ant}=50\Omega$ at each iteration until the impedance mismatch error is minimized. The matching network is also optimized intermittently with the load resistor. For a given channel realization, the waveform weights are designed and are then used to generate in Matlab the waveform $y(t)$ as in (4). Quantity $y(t)$ is then fed into the PSpice circuit simulator to generate the voltage source $\textnormal{Vs}=v_s(t)=2 y(t) \sqrt{R_{ant}}$ in Fig. \ref{circuit}. The DC power collected over the load can then be measured.

\par We evaluate the performance of WPT waveforms in a point-to-point scenario representative of a WiFi-like environment at a center frequency of 5.18GHz with a 36dBm transmit power, isotropic transmit antennas, 2dBi receive antenna gain and 58dB path loss in a large open space environment with a NLoS channel power delay profile obtained from model B \cite{Medbo:1998b}. Taps are modeled as i.i.d.\ CSCG random variables and normalized such that the average received power is $-20$dBm, i.e. $10\mu$W. The frequency gap is fixed as $\Delta_f=B/N$ and $B=1$MHz. The $N$ sinewaves are centered around 5.18GHz.

\begin{figure}
\centerline{\includegraphics[width=\columnwidth]{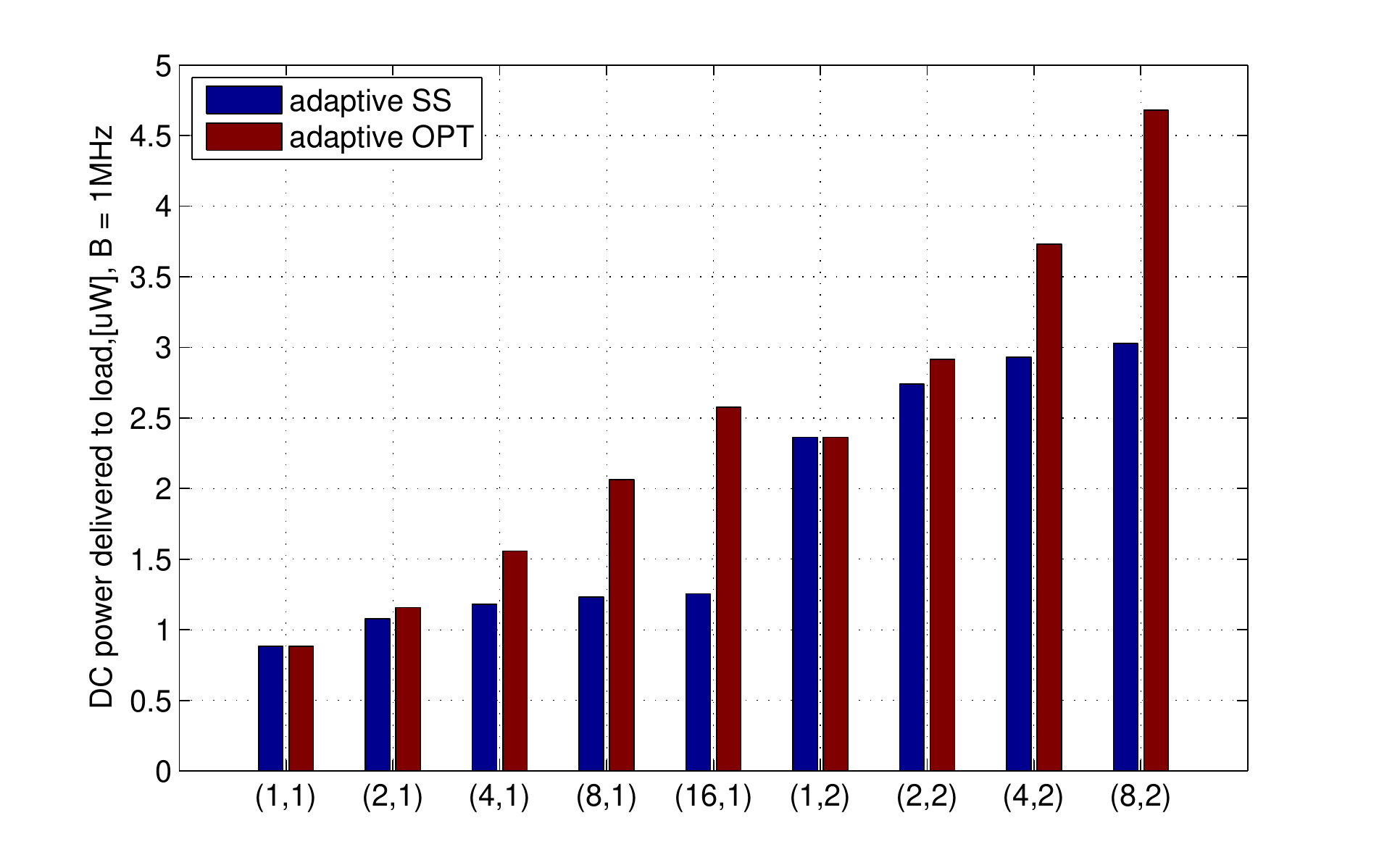}}
  \caption{Average DC power as a function of $(N,M_t)$ with $B=1$MHz.}
  \label{pspice_results_1MHz}
\end{figure}

\par Fig. \ref{pspice_results_1MHz} displays the average harvested DC output power for $B=1 $MHz with two waveforms, namely Adaptive Single Sinewave (SS) which is optimal for a 2nd-order (linear) term maximization (allocating all power to a single sinewave, i.e.\ the one corresponding to the strongest channel) and Adaptive OPT which solves problem \eqref{P2} with $n_o=4$ (and therefore accounts for non-linearity) using the reverse GP approach. In other words, Adaptive SS maximizes $e_2$ and assumes $e_3$ is constant. Adaptive OPT on the other hand optimizes the output DC power for a given transmit RF power and therefore maximizes the entire link efficiency $e_2 \times e_3$. We note the significant (and increasing as $N$ grows for any $M_t$) gains achieved by the nonlinear model-based design over the linear model-based counterpart. This shows that the non-linearity of the rectifier is clearly not negligible and the linear model is not sufficient to characterize correctly the rectenna behaviour. The sharp increase with $N$ of the DC power with the Adaptive OPT waveform cannot be explained based on the linear model, since according to the linear model the Adaptive SS waveform should be the one leading to the highest DC power. This shows that maximizing $e_2$ by assuming a constant $e_3$ is inefficient for most realistic input power and $N$ and confirms the observations made in Fig. \ref{2nd_vs_4th}.

\subsection{Extension and Future Work}
\par Previous discussions open the door to many important and exciting research avenues left for future work. Some of them are highlighted below.
\par This section highlighted the importance of understanding and modeling the wireless power channel (concatenation of the propagation channel and rectenna) and formulating a complete link optimization (transmitter to rectenna output) in order to design an efficient WPT architecture. Similarly to channel modeling in wireless communications where various types of models exist either for system analysis/design or performance evaluations, the design of wireless power-based system calls for various types of wireless power channel models. Models used for system design need to be sufficiently accurate but tractable enough in order to conduct link and system optimization. This section dealt with a key property of the rectifier, namely its non-linearity but other properties may need to be captured in the model, such as impedance and input power mismatch, harmonics, presence of multiple diodes in the rectifier, etc. Some results on the input power mismatch and its implication on system design can be found in \cite{Boshkovska:2015}.
\par A fundamental question that is arising from the waveform design in this section is, given a spectrum bandwidth $B$, what is the best/optimal way to transmit energy so as to maximize the output DC power? Answering this question would help understanding how to make the best use of the RF spectrum for WPT and lay a fundamental underlying theory of WPT.
\par A somewhat related question is whether WPT waveforms should be deterministic or modulated and whether modulated waveforms incur some loss with respect to deterministic waveforms. Answering those questions would help understanding the key tradeoff between transmitting information and power and helps designing unified wireless power and communications systems.
\par The fundamental role played by CSI in WPT remains largely unknown. The CSI acquisition/feedback in WPT also remains a serious challenge. Some interesting ideas along this line have been discussed in Section~\ref{sec:channelEstimation}. However, those approaches rely on the linear model. It is unclear yet whether a similar approach can be used over the non-linear wireless power channel.
\par Practical implementation of WPT requires low-complextiy algorithm design and the techniques involved to solve the optimization problem are not implementation-friendly. This calls for low-complexity approaches whose performance comes very close to the optimal design of the reverse GP. This would avoid solving computationally intensive optimization problems and would be much more suitable for practical implementation. Some ideas along those lines can be found in \cite{Huang:2016}, \cite{Huang:2016b}, \cite{Clerckx:2016d}.
\par WPT is the fundamental building block of various types of wireless powered systems (e.g.\ WPT, SWIPT, wireless powered communications, backscatter communications), which motivates a bottom-up approach where any wireless powered system is based on an established theory and design of the underlying WPT. The waveform design and the rectifier non-linearity tackled in this section have direct consequences on the design of SWIPT, wireless powered communications and backscatter communications. For instance, some preliminary results on SWIPT waveforms have been reported in \cite{Clerckx:2016c}, where it is shown that the superposition of multisine and OFDM waveforms enlarges the rate-energy region compared to an OFDM-only transmission. This originates from the non-linearity of the rectifier and the fact that the OFDM waveform, due to the randomness of the information, is less efficient than a (deterministic) multisine waveform to convert RF power to DC power. Assuming zero-mean Gaussian input distribution for the OFDM waveform, the superposition with the deterministic component of the multisine creates a non-zero mean Gaussian input distribution which is shown to outperform the conventional capacity-achieving zero-mean Gaussian input distribution in terms of rate-energy trade-off. More research endeavors are required to further investigate along this direction.


\section{Further Discussions}\label{sec:FurtherDiscussion}
In this section, we provide further discussions on various other issues pertaining to WPT.
\subsection{Safety and Health Issues}\label{sec:Safety}
Like any other RF-based wireless systems, WPT systems need to comply with the various safety guidelines to minimize the potential biological effects caused by RF energy \cite{815},\cite{817}. Though no existing studies show a clear evidence between electromagnetic radiation and health impairments, it has been well known that high level RF exposure is harmful to human body due to the rapid heating and thus possibly causes damage to biological tissue \cite{814},\cite{816}. Two widely adopted measures on RF exposure are {\it specific absorbtion rate} (SAR) and {\it maximum permissible exposure} (MPE) \cite{815}, which could be taken into account for WPT systems design.

SAR is a measure of the rate at which energy is absorbed by the human body when exposed to RF field. It has the units of watts per kilogram (W/kg), with the value typically obtained via experimental measurement averaged over a small sample of tissue (typically 1g or 10g of tissue). SAR is commonly used for testing the portable wireless devices that need to be used less than 20cm from the human body. For instance, FCC requires that all handphones sold in United States should not exceed the SAR level 1.6W/kg for partial body exposure \cite{818}, and similar SAR limits are specified by other countries. Note that SAR value not only depends on the source transmission power, but also on how the power is distributed over the tissue under test. Recently, it has been shown that for multi-antenna systems, the resulting SAR value can be modeled as a quadratic form of the transmitted signal as \cite{819}, \cite{820}
\begin{align}
\mathrm{SAR}=\mathbb{E}\left[\mathbf x^H \mathbf R \mathbf x \right]= \tr(\mathbf R \mathbf S),
\end{align}
where $\mathbf x$ is the transmitted signal vector with covariance matrix $\mathbf S$, and $\mathbf R$ is the SAR matrix depending on the SAR measurement setup such as the geometry and part of the body that is being tested. The generalized constraints with multiple SAR limitations can then be modeled as
\begin{align}
\tr(\mathbf R_g \mathbf S)\leq \eta_g, \ g=1,\cdots, G, \label{eq:SAR}
\end{align}
where $G$ is the total number of SAR constraints, $\mathbf R_g$ is the $g$-th SAR matrix, $\eta_g$ is the $g$-th constrained value. Note that \eqref{eq:SAR} resembles similar power constraints on wireless communication systems, such as as the interference-temperature constraints in cognitive radio systems \cite{331}, \cite{312}.

Another commonly used RF exposure limit is MPE, which is defined as the highest level of RF exposure to which a person may be exposed without incurring an established adverse health effect \cite{815}. MPE is usually expressed as power densities in W/m$^2$. For instance, IEEE has specified that the MPE limits from $1.5$GHz to $100$GHz for the general public or uncontrolled exposure is $10$ W/m$^2$ \cite{815}. Unlike SAR, MPE is generally a calculated quantity based on the source transmission power, antenna gain, propagation distance, etc. For multi-antenna WPT systems with highly directional transmissions, there exist highly localized areas or {\it RF hot spots}, usually along the beamforming directions, with much higher RF intensity than other areas, as shown in Fig.~\ref{F:PowerDisComparison}(a). WPT systems could be designed to ensure that the RF intensities at all locations (including the RF hot spot) do not exceed the MPE limit \cite{821}. Alternatively, for WPT systems with relatively high-power requirement, it is more feasible to guarantee the MPE limit only at non-intended serving   areas. In this case, additional measures (such as building physical fence) must be taken to prevent people from entering into the RF hot spot. Furthermore, it is interesting to note that the two channel reciprocity based designs, i.e., reverse-link training and the retrodirective amplification schemes presented in Section~\ref{sec:Pt2Pt}, have the inherent safety mechanism. Specifically, if a particular path between the ET and ER is blocked by a person, then the reverse-link pilot signal sent by the ER will not arrive at the ET through the blocked path. As a result, no energy beamforming will be formed towards the person during the forward energy transmission phase. For such setups, the more frequently the reverse training is performed, the shorter time an intruder will be exposed to the energy beamforming directions with potentially high energy intensity, but at the cost of increased training overhead. This thus requires a design trade-off between maximizing WPT performance and minimizing safety risk, which deserves more in-depth investigations. Besides, by comparing the two plots in Fig.~\ref{F:PowerDisComparison}, it is found that the distributed antenna system is more appealing from the safety perspective, since it avoids the hot spot issue along the energy beamforming directions as in the co-located antenna system.

\subsection{Massive MIMO and MmWave WPT}
Massive MIMO is a key enabling technology for the fifth-generation (5G) wireless communication systems 
 by tremendously increasing the spectrum efficiency via deploying a large number of antennas (say, hundreds or even more)  at the BSs \cite{373}, \cite{374,375,497}. For WPT systems, massive MIMO is also an appealing technique to enhance the end-to-end power transfer efficiency by deploying large antenna arrays at the ETs \cite{Huang:2016b}, \cite{Huang:2016}, \cite{518,845,844,843}. Intuitively, with perfect CSI at the ET, the energy beamforming gain, and hence the end-to-end power transfer efficiency, increases linearly with the number of antennas $M_t$ at the ET.  Remarkably, it has been shown that even with reverse-link based channel training as discussed in Section~\ref{sec:Pt2Pt},  the net harvested energy at the ER also increases linearly with $M_t$ as $M_t\rightarrow \infty$ \cite{528},\cite{558}. As compared to massive MIMO communications, massive MIMO WPT systems possess several new characteristics. In particular, the {\it pilot contamination issue}, where the users in neighboring cells severely interfere with each other due to the sharing of pilot sequences for channel estimation, is regarded as a main performance bottleneck for massive MIMO communications \cite{373}. In contrast, pilot contamination could even be beneficial  for WPT systems \cite{844}, since the power directed towards the non-intended directions due to pilot contamination can also be harvested by other ERs, instead of causing the detrimental interference as in communication systems. Furthermore, contrary to other works, in \cite{Huang:2016b}, \cite{Huang:2016}, massive MIMO WPT was studied in light of the non-linear energy harvesting model. Waveform strategies (accounting for energy beamforming) suitable for a large scale multi-antenna multi-sine WPT architecture were derived in single-user and multi-user scenarios. The benefits of exploiting the non-linearity of the rectifier in the system design was confirmed for Massive MIMO WPT.

Another promising technology for 5G  is millimeter wave (mmWave) communication \cite{566},\cite{569}, which utilizes the large available bandwidth at mmWave frequencies (typically from around 30GHz to 300GHz) and large antenna arrays at the BSs (also possibly at the mobile stations) to enable Giga-bits per second (Gbps) radio access. Thanks to the significantly reduced signal wavelength, large antenna arrays for mmWave systems can be packed compactly with small form factors, which makes mmWave also an appealing technique for WPT applications \cite{833}. However, mmWave signals usually suffer from poor penetration and diffraction capabilities, which make them sensitive to blockages. Thus, more research efforts are needed to develop effective techniques to realize reliable mmWave WPT systems.

Both massive MIMO and mmWave WPT systems rely on large antenna arrays at the ETs to achieve highly directional transmissions. This renders the traditional digital signal processing technique that requires one RF chain for each antenna costly, in terms of both hardware implementation and power consumption. Extensive efforts have been recently devoted to enabling cost-effective massive MIMO and mmWave wireless communications via techniques such as analog beamforming \cite{574}, hybrid analog/digital processing \cite{578}, or the advanced lens antenna arrays \cite{485},\cite{823}. The extension of such techniques for cost-effective WPT is a promising avenue for future research  \cite{846}.


\subsection{Wireless Charging Control}\label{sec:ChargingControl}
For WPT networks with a large number of ETs serving massive ERs, effective wireless charging control mechanisms need to be devised for real-time decisions such as user scheduling \cite{813}, frequency usage, ER and ET association and on/off control \cite{868}, and the amount of power to be transmitted, etc. Efficient wireless charging control is in general a complicated task that in general needs to both minimize  energy outage and also avoid  battery over-charging/overflow, which depends on various factors including the CSI between the ETs and ERs, the ERs' battery status information (BSI) and their energy demands \cite{861}, user fairness, etc. For example, from the perspective of maximizing the overall energy transfer efficiency, the ERs that have the best channel conditions should be scheduled for WPT at each time slot. On the other hand, to avoid battery depletion of the nodes and hence prolong the network lifetime, higher priority should be given to those nodes with low residual battery energy and high energy consumption demands. Moreover, to reduce control and feedback overhead, wireless charging control decisions usually need to be made in a distributed manner based on local information. 
 For WPT systems over wide-band frequency-selective channels,  the authors in \cite{835} proposed a voting-based distributed charging control protocol. With this protocol, each ER estimates the wide-band channels, casts the votes for some strong sub-channels for energy transmission and sends them to the ETs along with its BSI, based on which the ETs  allocate their transmit power over the sub-channels without the need of centralized control.





\subsection{Joint Design with Wireless Communications}
As  wireless power and communication systems both use RF waveforms as the energy/information carrier, they could be jointly designed to seamlessly integrate each other. There are three main lines of research along this direction, namely {\it SWIPT, wireless powered communications, and coexisting design of WPT and wireless communication systems}.

In SWIPT systems \cite{503,504,478}, information and power are transmitted concurrently from the same nodes using the same RF waveforms, where the information and energy receivers could either be co-located or separated.  Numerous research efforts have been recently devoted to maximizing the achievable {\it rate-energy} region under various setups of  SWIPT systems \cite{745,746,747,534}. At the transmitter side, the power allocation, beamforming/precoding, waveform, frequency selection, etc., need to be carefully optimized to achieve the optimal trade-off between information and power transmissions  \cite{Clerckx:2016c}, \cite{527,526,847,848,872}. At the receiver side, various information/energy receiving strategies have been proposed. For example, for co-located energy/information receivers, {\it time switching} and {\it power splitting} are two prominent strategies to achieve both energy and information receptions at the same node \cite{478}, \cite{521,513,862}. The performance benefits of time switching vs power switching nevertheless highly depends on the rectifier (linear vs non-linear) model \cite{Clerckx:2016c}.  An {\it integrated receiver} architecture has also been proposed \cite{514}, where the information is encoded in the energy signal by varying its power levels over time for achieving continuous information transfer without degrading the power transfer efficiency. Researchers are now working actively to practically realize the promising concept of SWIPT \cite{734}. Another interesting application in SWIPT systems is to utilize the energy signals as artificial noise to protect the messages for the information receivers from being eavesdropped by the non-intended energy receivers \cite{863}, \cite{864}.

For wireless powered communication systems \cite{515,525,744,753,742,743}, \cite{516,530,873,874}, the energy for wireless communications at the devices is obtained  via WPT upon usage. In this case, both the wireless power and communication links need to be jointly designed, under the new constraints that the harvested power at the  wireless devices should be no smaller than that used for communications. Wireless powered communication systems are commonly studied based on the {\it harvest-then-transmit} (HTT) protocol \cite{515},\cite{836}, where the wireless devices first harvest sufficient energy with WPT for certain time duration before initiating information transmission. Alternatively, RF energy harvesting and wireless communication could occur concurrently at the wireless devices with the novel concept of {\it energy/information full-duplex}, where each wireless device performs {\it simultaneous energy harvesting} and {\it information transmission}, with the additional benefit of self-energy recycling \cite{551}. For multi-user wireless powered communication systems, a ``doubly near-far'' problem has been revealed, where a far ER from the ET suffers from higher loss than a near ER for both downlink wireless power transfer and uplink information transmission \cite{515}.  Various techniques have been proposed to mitigate the doubly near-far problems, such as via user cooperation \cite{541} or separating the energy and information access points \cite{742}.

Last, WPT and wireless communication systems could also be designed to operate separately, but with the coexisting issues properly addressed. Besides using orthogonal bands for information and power transmissions, the two systems could share the same band for more efficient spectrum utilization, as long as the interference to wireless communications caused by energy signals is effectively mitigated \cite{822}, \cite{837}. Note that different from that in wireless communication systems, the presence of the coexisting information signal actually contributes as the additional RF  source to energy harvesting at the WPT receivers. Cognitive radio techniques have been recently applied for coexisting wireless information and power transfer systems \cite{510}, \cite{849}. Furthermore,  it is shown in \cite{822} that the single-beam time sharing scheme discussed in Section~\ref{sec:PowerRegion} only occupies one spatial dimension at each time interval, and thus is preferred for coexisting wireless power and communication systems.
Another major challenge for the coexisting systems stems from the fact that the energy signals are usually orders-of-magnitude stronger than the information signals due to the different receiver sensitivities for information decoding and energy harvesting. This may cause severe signal distortion at the information receiver due to device saturation (e.g., power amplifiers and analog-to-digital converters). There are some existing studies to address this issue, e.g., by using lens antenna arrays to automatically separate the energy and information signals in the antenna domain \cite{838}, or by  suppressing the strong energy signal prior to digital signal processing using analog circuits at the receiver frontend \cite{839}.

\section{Conclusions}\label{sec:Conclusion}
This article provides a tutorial overview on the main communication and signal processing techniques for WPT systems. Under the linear energy harvesting model, the main techniques for enhancing the RF power transfer efficiency are firstly discussed in single-user WPT systems, including energy beamforming,  channel acquisition, retrodirective amplification, etc. For multi-user WPT systems, the various networking architectures with different levels of cooperation among the ETs are then introduced, followed by the power region characterizations via convex optimization techniques. The nonlinear energy harvesting model and the corresponding waveform optimizations to further enhance the power transfer efficiency are presented next. Finally, we provide further discussions on various other topics pertaining to WPT, including safety and health issues, WPT using massive MIMO and mmWave techniques, wireless charging control, and wireless power and communication systems co-design. It is hoped that the techniques presented in this article will help inspiring future researches in this exciting area as well as  paving the way for practically designing and implementing efficient  WPT systems in the future.

\bibliographystyle{IEEEtran}
\bibliography{IEEEabrv,IEEEfull}

\end{document}